\documentclass[pra,twocolumn,showpacs,superscriptaddress,nofootinbib]{revtex4-1}
\usepackage{psfrag,graphicx}
\usepackage{amsfonts,amssymb,amsmath}      
\usepackage{graphicx}
\usepackage{dcolumn}
\usepackage{bm}
\usepackage{float}
\usepackage{hyperref}
\usepackage{accents}
\usepackage{bbding}
\usepackage{color,soul}
\usepackage{epstopdf}

\newcommand{\ellipse}{\raisebox{-1pt}{\scalebox{1.3}[.4]{$\circ$}}}
\newcommand{\halo}{\accentset{\ellipse}}
\newcommand{\erf}[1]{Eq.~(\ref{#1})}
\newcommand{\beq}{\begin{equation}}
\newcommand{\eeq}{\end{equation}}
\newcommand{\nn}{\nonumber}

\newcommand{\erfs}[2]{Eqs.~(\ref{#1})--(\ref{#2})}

\newcommand{\dg}{^\dagger}
\newcommand{\ddg}{^\ddagger}

\newcommand{\sch}{Schr\"odinger}

\newcommand{\Tr}{\text{Tr}}

\newcommand{\tp}{^{\top}}

\newcommand{\s}[1]{\hat{\sigma}_{#1}}
\renewcommand{\c}{_{\text{C}}}
\newcommand{\ob}{_{\text{o}}}
\newcommand{\un}{_{\text{u}}}
\newcommand{\m}{_{\text{m}}}
\newcommand{\p}{_{\text{p}}}
\newcommand{\ex}[1]{\langle{#1}\rangle}
\newcommand{\dd}{{\rm d}}
\newcommand{\dwn}{_{\downarrow}}
\newcommand{\up}{_{\uparrow}}
\newcommand{\ddt}[1]{\frac{\dd{#1}}{\dd t}}
\newcommand{\hL}{\halo\Lambda}

\newcommand{\SHUR}{\sch-Heisenberg uncertainty relation}

\newcommand{\past}[1]{\overleftarrow{#1}}
\newcommand{\fut}[1]{\overrightarrow{#1}}
\newcommand{\both}[1]{\overleftrightarrow{#1}}
\newcommand{\fil}{_{\text F}}
\newcommand{\rfil}{_{\text R}}
\newcommand{\sm}{_{\text S}}
\newcommand{\swv}{_{\rm SWV}}
\newcommand{\swd}{_{\rm \red SWD}}
\newcommand{\god}{_{\text T}}
\newcommand{\inv}{^{-1}}
\newcommand{\bx}{{\bf x}}
\newcommand{\bcx}{{\check{ \bf x}}}
\newcommand{\by}{{\bf y}}
\newcommand{\bv}{{\bf v}}
\newcommand{\bw}{{\bf w}}

\newcommand{\hV}{\halo{V}}

\newcommand{\xfil}{\ex{\bx}\fil}

\newcommand{\xsm}{\ex{\hat \bx}\sm}
\newcommand{\xgod}{\ex{\hat \bx}\god}
\newcommand{\K}{{\cal K}}

\definecolor{nblue}{rgb}{0.06,0.3,0.73}
\definecolor{nblack}{rgb}{0,0,0}
\definecolor{nred}{rgb}{0.9,0.1,0.1}
\definecolor{nmagenta}{rgb}{0.7,0.0,0.3}

\newcommand{\red}{\color{nred}}

\newcommand{\blk}{\color{nblack}}

\newcommand{\hbx}{\hat\bx}

\begin{document}

\title{Linear Gaussian Quantum State Smoothing: Understanding\\ the optimal unravelings for Alice to estimate Bob's state}

\author{Kiarn T. Laverick}
\affiliation{Centre for Quantum Computation and Communication Technology 
(Australian Research Council), \\ Centre for Quantum Dynamics, Griffith University, Nathan, Queensland 
4111, Australia\\}
\author{Areeya Chantasri}
\affiliation{Centre for Quantum Computation and Communication Technology 
(Australian Research Council), \\ Centre for Quantum Dynamics, Griffith University, Nathan, Queensland 
4111, Australia\\}
\affiliation{Optical and Quantum Physics Laboratory, Department of Physics, Faculty of Science, Mahidol 
University, Bangkok 10140 Thailand}
\author{Howard M. Wiseman}
\affiliation{Centre for Quantum Computation and Communication Technology 
(Australian Research Council), \\ Centre for Quantum Dynamics, Griffith University, Nathan, Queensland 
4111, Australia\\}

\date{\today}
\begin{abstract}
Quantum state smoothing is a technique to construct an estimate of the quantum state at a particular time, 
conditioned on a measurement record from both before and after that time. 
The technique assumes that an observer, Alice, monitors part of the environment of a quantum system and 
that the remaining part of the environment, unobserved by Alice, is measured by a secondary observer, Bob, 
who may have a choice in how he monitors it. The effect of Bob's measurement choice on the effectiveness 
of Alice's smoothing has been studied in a number of recent papers. Here we expand upon the Letter which 
introduced linear Gaussian quantum (LGQ) state smoothing [Phys. Rev. Lett., {\bf 122}, 190402 (2019)]. 
In the current paper we provide a more 
detailed derivation of the LGQ smoothing equations and address an open question about Bob's optimal 
measurement strategy. Specifically, we develop a simple hypothesis that allows one to approximate the 
optimal measurement choice for Bob given Alice's measurement choice. By `optimal choice' we mean the 
choice for Bob that will maximize the purity improvement of Alice's smoothed state compared to her 
filtered state (an estimated state based only on Alice's past measurement record). The 
hypothesis, that Bob should choose his measurement so that he observes the back-action on the system 
from Alice's measurement, seems contrary to one's intuition about quantum state smoothing. Nevertheless 
we show that it works even beyond a linear Gaussian setting.
\end{abstract}
\maketitle

\section{Introduction}
In parameter estimation, the task is to estimate unknown parameters, denoted by a vector $\bx$, 
from available information such as measurement records. A powerful tool for parameter estimation is the 
probability density function (PDF), 
often called the {\em state} of the system, as it is possible to compute from this any estimate of $\bx$, 
e.g., the mean or the mode of the PDF. This turns the problem into one of state estimation. There are 
numerous techniques for classical state estimation. 
Specifically, for continuous measurements, there are the techniques of {\em filtering} and {\em smoothing}
\cite{Weinert01,Hay01,BroHwa12,Ein12,Fri12,vanTrees1} for classical states. Filtering uses any 
measurement information 
prior to the estimation time $\tau$, the `past' measurement record $\past{\blk\rm O}$, to estimate the state of 
the system, yielding the {\em filtered} state $\wp\fil(\bx) := \wp(\bx|\past{\blk\rm O})$. The complement to 
the filtered state is the {\em retrofiltered} effect $E\rfil(\bx) := \wp(\fut{\blk\rm O}|\bx)$, more commonly referred 
to as the likelihood function \cite{Ein12,BroHwa12,Sarkka13} for the future measurement record 
$\fut{\blk\rm O}$ given $\bx$. The estimation technique of smoothing combines the filtered state and 
retrofiltered effect to obtain a {\em smoothed} state $\wp\sm(\bx) := \wp(\bx|\both{\blk\rm O}) \propto 
E\rfil(\bx)\wp\fil(\bx)$, conditioned on both past and future measurement records, the `past-future' 
measurement record $\both{\blk\rm O}$. While smoothing may be inapplicable for some purposes, as it 
requires information after the estimation time, it is a more accurate estimation technique for data 
post-processing than filtering as it utilises more information.

As we make the transition to quantum technologies, it becomes increasingly important to estimate the 
quantum state $\rho$ of a system. There are well-known techniques to estimate quantum state preparation 
from an ensemble of measurement results, e.g.,~tomography \cite{DarParSac03}. Here, however, we 
are interested in 
techniques using a single realization of a continuous measurement record, such as quantum trajectory 
theory \cite{Belav87,Bel92,WisMil10}. This technique is analogous to the 
classical technique of filtering in that it only uses the past measurement 
record to obtain the filtered quantum state $\rho\fil(\tau)$. 
As in the classical case, the complement of the filtered quantum state is the retrofiltered quantum effect
$\hat E\rfil(\tau)$, a positive operator defined such that $\Tr[\hat{E}\rfil\rho] = \wp(\fut{\blk\rm O}|\rho)$.

For the quantum analog of smoothing, it is not as simple as combining the filtered state and the 
retrofiltered effect as it was in the classical case. If we were to combine them following the pattern of the 
classical case,~$\varrho(\tau) \propto \rho\fil(\tau)\hat{E}\rfil(\tau)$, the 
resulting operator would not be a valid quantum state. That is in general, the operator is
not positive semidefinite \cite{Tsa09a,Tsa09b,GJM13,GueWis15,Ohki15,Tsa19,LCW-QS19}. 
We do not want to give the reader the impression that this operator is useless; in fact, it has an interesting 
connection to weak values \cite{ABL64,AAV88,Tsa09b}. Consequently, a symmetrized 
version of $\varrho(\tau)$ has been referred to as the smoothed weak-value (SWV) state $\varrho\swv$ 
\cite{LCW19}. {\red A different way of doing quantum smoothing, also introduced by Tsang \cite{Tsa09b}, is to multiply together the Wigner distributions of $\rho_{\rm F}(\tau)$ and $\hat{E}_{\rm R}(\tau)$, and normalize, to get a smoothed Wigner distribution (SWD) $W_{\rm SWD}$. The ``state'' corresponding to $W_{\rm SWD}$ is also, in general, not a valid quantum state.}

\begin{figure}
\includegraphics[scale=0.335]{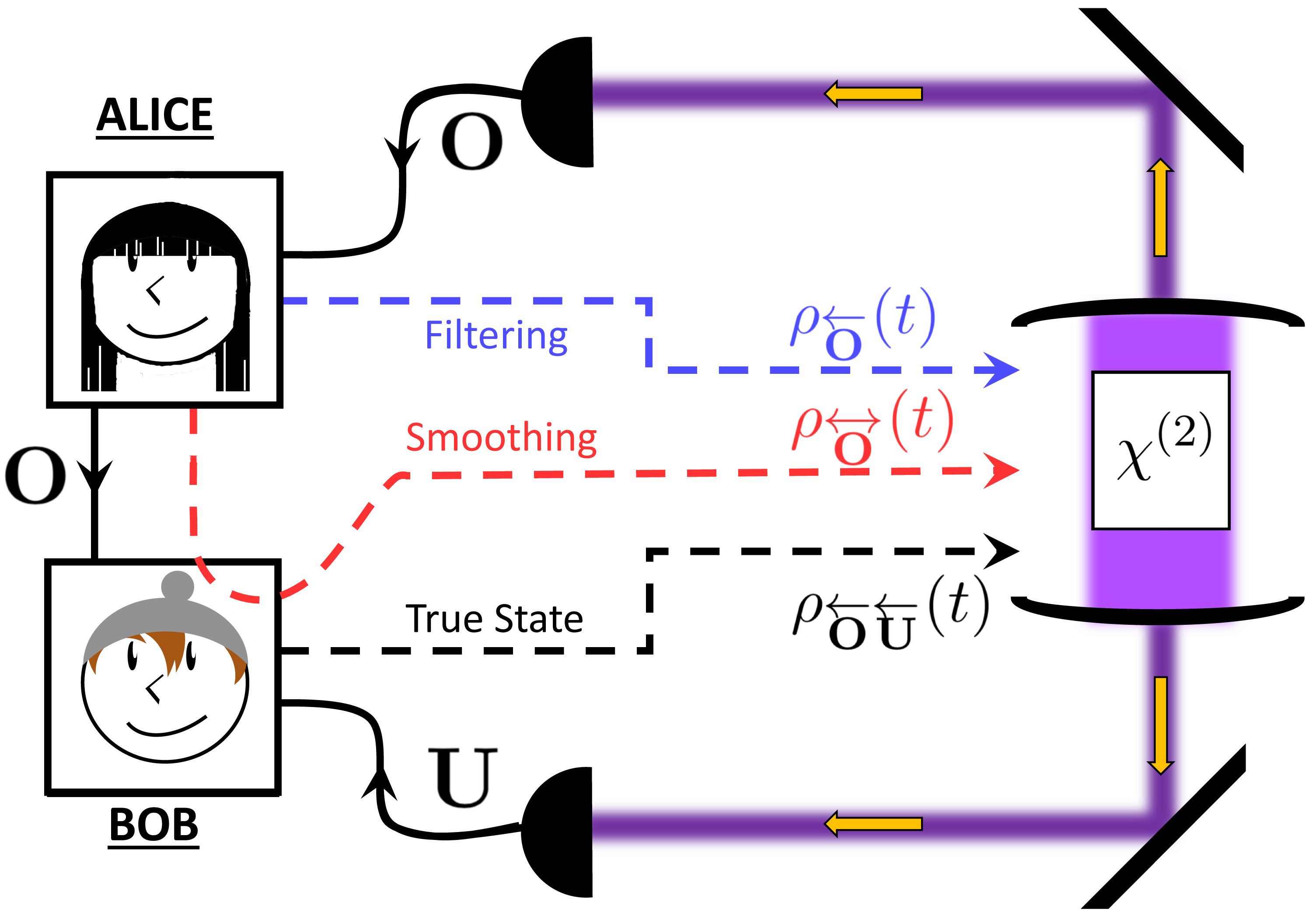}
\caption{A diagrammatic representation of the quantum state smoothing formalism.  
Bob, who has access to both the observed record ${\blk\rm O}$ and the unobserved record ${\blk\rm U}$, is
able to obtain the best estimate of the quantum state, the true state 
$\rho\god :=\rho_{\protect\past{\blk\rm O}\protect\past{\blk\rm U}}$ {\blk of the quantum system ${\cal Q}$}.
Alice, on the other hand has access to only the observed record ${\blk\rm O}$. If Alice does not know of the 
existence of the ${\blk\rm U}$, then her best estimate would be the filtered estimate 
$\rho\fil :=\rho_{\protect\past{\blk\rm O}}$. 
However, if Alice knows the measurement setting Bob used to obtain ${\blk\rm U}$, she 
can utilise the full past-future observed record to obtain the smoothed state 
$\rho\sm:=\rho_{\protect\both{\blk\rm O}}$, which is a more 
accurate estimate of Bob's true state than the filtered state.}
\label{Fig-QSS}
\end{figure}

There is, however, a quantum state smoothing formalism developed by 
Guevara and Wiseman \cite{GueWis15} which guarantees a valid smoothed quantum state. The formalism 
considers a quantum system partially observed by an observer, Alice, whose task is to estimate the true 
state of the systems using only her observed record. However, for Alice to obtain a 
valid smoothed quantum state, that is, a state conditioned on her past-future measurement record, 
it is necessary to introduce a secondary observer, say Bob, who gathers all information 
unobserved by Alice, see Fig.~\ref{Fig-QSS}. By using both Alice's and Bob's measurement records to estimate 
the quantum state, we would obtain the {\em true} quantum state, a state containing maximal information about 
the quantum system. The true state is crucial to calculating the smoothed state.

The smoothed quantum state has been shown to offer a better estimate of the true state than the 
conventional filtered state, where the improvement is quantified by the state purity 
\cite{GueWis15,LCW19,CGW19}. Interestingly, the purity improvement of the smoothed state over the filtered 
state depends on both Alice's and Bob's choices of measurement on their parts of the system's environment. 
Note, these choices do not affect the unconditioned system evolution, described by a master equation. 
This raises an interesting question: 
How should Alice observe and `unobserve' (that is, Bob observe) the quantum system in order to obtain 
the maximum purity improvement for the smoothed quantum state? 
Recently \cite{CGW19}, the optimal measurement strategy for Alice and Bob has been investigated for a 
single qubit example. However, due to the vast number of unobserved measurement records that are 
needed in order to calculate the smoothed quantum state in such a system, the authors were only able to 
consider a handful of measurement scenarios. 

Since the original proposal in 2015 \cite{GueWis15}, the quantum state smoothing theory has been adapted 
by the present authors to linear Gaussian quantum (LGQ) systems \cite{LCW19}. Thanks to the nice 
properties of LGQ systems, the theory of Ref.~\cite{LCW19} provided simple closed-form solutions for the 
smoothed quantum state, enabling its properties to be investigated either analytically or semianalytically 
\cite{LCW19,LCW-QS19}. If we restrict our analysis to LGQ systems, though we are also restricting to 
diffusive-type unravelings of the system, we can drastically increase the number of measurement 
scenarios for Alice and Bob in the search for the optimal measurement strategy.
As a result, we can numerically determine the optimal diffusive measurement scenario for Alice 
and Bob for any type of LGQ system. But can we understand the results intuitively?

In this paper, we first review the 
necessary theory required for LGQ state smoothing, and provide a more detailed derivation of the theory 
than that presented in Ref.~\cite{LCW19}. We then present numerically simulated LGQ trajectories, 
showing their means and covariances, of the filtered, \red SWD \blk, and smoothed quantum states. This is to 
observe the differences in these estimators and analyze their properties as a function of time. As expected, 
we observe that the smoothed quantum state estimates the true state better than the filtered state could. The 
\red SWD \blk state, on the other hand, performs very differently.

As the main focus of this paper, we present three possible hypotheses for the optimal measurement strategy 
for Alice and Bob, and study how well they predict the optimal measurements found 
numerically for two LGQ physical systems: an on-threshold optical parametric oscillator and a stochastic
linear attenuator. The most successful strategy has a surprisingly counter intuitive logic to it. Lastly, we 
generalize the logic behind the most successful hypotheses from the LGQ setting to the qubit setting by 
defining analogous quantities for a driven qubit measured using homodyne detections. Moreover, we find 
that the success of the counter intuitive strategy is replicated in the qubit system.
 
The structure of this paper is as follows. In Sec.~\ref{sec-LGC} we will briefly review the classical linear 
Gaussian (LG) state estimation. Then, in Sec.~\ref{sec-LGQ} we review LGQ systems along with the 
LGQ state smoothing theory. Next, in Sec.~\ref{sec-PS} we introduce the two physical systems that we 
will consider throughout the paper. 
We simulate the trajectories for the filtered, true, \red SWD \blk, and 
smoothed quantum states in Sec.~\ref{sec-Traj}. 
Finally, in Sec.~\ref{sec-Opt} we find a simple hypothesis for the best measurement strategy for Alice and 
Bob to maximize the purity of the smoothed state compared to the filtered state, which works for our two 
LGQ examples and, suitably generalized, for a very different qubit example. 

\section{Classical LG State Estimation}\label{sec-LGC}
For a classical dynamical system, a state of knowledge of the system is defined as the PDF 
$\wp({\blk\bcx})$, where ${\blk \bcx = (\check{x}_1,\check{x}_2,...,\check{x}_D)\!\tp}$ is the vector of $D$ 
parameters required to completely describe the system, with $\top$ denoting the transpose. {\blk Note, we have 
used the wedge mark on ${\blk \bcx}$ to make it clear that this is a dummy variable for the PDF and not the 
corresponding random variable which we denote by $\bx$.} We will restrict our analysis to Gaussian states, 
$\wp(\bcx) = g(\bcx;\ex{\bx},V)$. That is, the state is specified by its mean $\ex{\bx}$ and covariance 
matrix $V = \ex{\bx\bx\tp} - \ex{\bx}\ex{\bx}\!\tp$. In order to guarantee that the state remains Gaussian 
throughout its evolution even when conditioned on continuous observation, the system must be initialized in 
a Gaussian state and must satisfy the following constraints 
\cite{WisMil10,Hay01,Weinert01,vanTrees1,BroHwa12,Ein12,Fri12}. First, the system's dynamical 
evolution must be described by a linear Langevin equation
\beq\label{LLE}
\dd\bx = A\bx\dd t + E\dd\bv\p\,.
\eeq
Here $A$ and $E$ are constant matrices and $\dd\bv\p$ is the process noise, which is a vector of 
independent Wiener increments that satisfies
\beq \label{WeinCond}
\mathbb{E}[{\rm d}\bv\p] = {\bf 0}\,, \qquad {\rm d}\bv\p({\rm d}\bv\p)\!\tp \!= I {\rm d}t\,,
\eeq
where $\mathbb{E}[...]$ denotes an ensemble average over all possible realisations of the noise. 
The second constraint is that any measurement record obtained must be linear in $\bx$, i.e.,
\beq
\by\dd t = C\bx\dd t + \dd\bv\m, 
\eeq
where $C$ is a constant matrix and $\dd\bv\m$ is the measurement noise, a vector of independent 
 increments satisfying similar conditions to \erf{WeinCond}. There may exist some correlations 
between the measurement noise and the process noise of the system, for example, from measurement 
back-action, which can be described by a cross-correlation matrix $\Gamma\tp\dd t = E\dd\bv\p(\dd\bv\m)\!\tp$ 
\cite{WisMil10}. We will note that the majority of classical texts 
\cite{Weinert01,Hay01,BroHwa12,Ein12,Fri12,vanTrees1} on this topic assume that $\Gamma = 0$.

The classical LG systems are defined by the above constraints. We can condition the 
estimate of the LG state on the past measurement record to obtain the filtered estimate 
$\wp\fil(\bcx) = g(\bcx,\xfil,V\fil)$, whose mean and covariance are given 
by the Kalman-Bucy filtering equations 
\cite{WisMil10,KaiFro68,Kai70,Kai73,BLP79}
\begin{align}
&{\rm d}\ex{\bx}\fil = \, A\ex{\bx}\fil {\rm d}t + {\cal K}^{+}[V\fil]{\rm d}\bw\fil\,,\label{cmf}\\
&\frac{{\rm d}V\fil}{{\rm d}t} = \, AV\fil +V\fil A\tp \!+ D - {\cal K}^{+} [V\fil] {\cal K}^{+} [V\fil]\tp\,,
\label{cVf}
\end{align}
with initial conditions $\ex{\bx}\fil(t_0) = \ex{\bx}_0$ and $V\fil (t_0) = V_0$.
Here, $\dd\bw\fil := \by\dd t - C\ex{\bx}\fil\dd t$ is a vector of innovations, $D = EE\tp$ is the diffusion 
matrix, and 
\beq \label{Kick}
\K^{\pm}[V] := VC\tp \! \pm \Gamma\tp
\eeq
is the optimal Kalman gain matrix, as a function of the covariance. 

As mentioned earlier, if we want to obtain a more accurate estimate of the state, we can utilise the 
past-future measurement record $\both{\blk\rm O}$ as opposed to the past record $\past{\blk\rm O}$ the 
filtered state 
uses. The smoothed state obtained by using $\both{\blk\rm O}$ can be calculated using the filtered state 
according to
\beq\label{Csm}
\wp\sm({\blk \bcx}) := \wp({\blk \bcx}|\both{\blk\rm O}) \propto E\rfil({\blk \bcx}) \wp\fil({\blk \bcx})\,,
\eeq
where we have assumed that the system is Markovian. 
To explicitly see the dependence on the measurement records, we remind the reader that the filtered state is 
a function of the past measurement record, $\wp\fil(\bcx):= \wp(\bcx|\past{\blk\rm O})$. The retrofiltered effect is 
the likelihood of a particular realization of a future measurement record occurring from a configuration 
${\blk\bcx}$, 
i.e., $E\rfil({\blk \bcx}):= \wp(\fut{\blk\rm O}|{\blk \bcx})$. Using Bayes' theorem \cite{Jaz07} results in \erf{Csm}. As 
we already 
have calculated the filtered state, all we need to calculate to obtain the smoothed state is the retrofiltered effect. 

If we apply Bayes' theorem to the retrofiltered effect, we obtain 
$E\rfil({\blk \bcx}) \propto \wp({\blk \bcx}|\fut{\blk\rm O})\wp(\fut{\blk\rm O})$. As we are using the retrofiltered effect 
to calculate the smoothed state, the future measurement record will be fixed and the probability 
$\wp(\fut{\blk\rm O})$ for that fixed record will be a constant. As a result, the retrofiltered effect is 
$E\rfil({\blk \bcx}) \propto \wp({\blk \bcx}|\fut{\blk\rm O})$, from which we can define a normalised retrofiltered 
effect $E\rfil'({\blk \bcx}) = \wp({\blk \bcx}|\fut{\blk\rm O})$. As we are limiting our discussion to Gaussian systems, 
the normalized retrofiltered effect will be a Gaussian, $E'\rfil({\blk \bcx}) = g({\blk \bcx};\ex{\bx}\rfil,V\rfil)$, where 
the retrofiltered mean $\ex{\bx}\rfil$ and corresponding covariance matrix $V\rfil$ are given by 
\begin{align}
&- {\rm d}\ex{\bx}\rfil =  -A\ex{\bx}\rfil {\rm d}t + {\cal K}^{-} [V\rfil]{\rm d}\bw\rfil,\label{crm}\\
&- \frac{{\rm d}V\rfil}{{\rm d}t} = -AV\rfil - V\rfil A\tp \!+ D - {\cal K}^{-} [V\rfil] {\cal K}^{-} [V\rfil]\tp.\label{cVr}
\end{align}
Here $\dd\bw\rfil= \by\dd t - C\ex{\bx}\rfil\dd t$ and $\K^-[V\rfil]$ is defined in \erf{Kick}. These retrofiltering 
equations evolve backwards in time, as evident from the negative sign on the left-hand side of both 
equations, from a final uninformative state with $V\rfil(T) = \infty$. However, due to the infinite final 
retrofiltered covariance, there is no sensible final condition for the retrofiltered mean. 

One can obtain more practical equations \cite{Fraser67}, which can be used in numerical computations and 
the upcoming \red SWD \blk state, by instead solving for the inverse 
retrofiltered covariance $\Lambda\rfil = V\rfil\inv$, referred to as an information matrix, and 
defining a new `informative' mean ${\bf z}\rfil = \Lambda\rfil \ex{\bx}\rfil$. Using the identity 
\beq\label{V2L}
\frac{\dd}{\dd t}V\inv = -V\inv \frac{\dd V}{\dd t} V\inv\,, 
\eeq
we obtain the equations for the retrofiltered informative mean and the information matrix
\begin{align}
&-\dd {\bf z}\rfil = (\tilde{A} - \tilde{D}\Lambda\rfil)\!\tp {\bf z}\rfil \dd t + (C\tp \!- \Lambda\rfil\Gamma\tp)
\by\dd t\,,\label{zr}\\
&- \frac{\dd \Lambda\rfil}{\dd t} = \Lambda\rfil\tilde{A} + \tilde{A}\tp \!\Lambda\rfil - \Lambda\rfil\tilde{D}
\Lambda\rfil + C\tp \!C\,,\label{Lr}
\end{align}
with $\tilde{A} = A - \Gamma\tp \!C$ and $\tilde{D} = D - \Gamma\tp\Gamma$. We can now simply set the 
final conditions to be ${\bf z}\rfil(T) = 0$ and $\Lambda\rfil(T) = 0$.

Finally, now that we have equations for both the filtered state and the retrofiltered effect, we can compute the 
smoothed state using \erf{Csm}. Due to the proportionalilty in \erf{Csm}, we can replace the retrofiltered 
effect $E\rfil({\blk \bcx})$ with its normalized counterpart $E\rfil'({\blk \bcx})$, as any proportionality constants will 
be accounted for during the normalization process. Since both the filtered state and retrofiltered effect are 
Gaussians, then by the multiplicative property of Gaussians, the smoothed state will also be Gaussian. That 
is, $\wp\sm({\blk \bcx}) = g({\blk \bcx};\ex{\bx\sm},V\sm)$, with smoothed mean and covariance 
\cite{Weinert01,Ein12,Sarkka13,Mayne66,Fraser67,FraPot69}
\begin{align}
&\ex{\bx}\sm = V\sm\left[V\fil\inv\ex{\bx}\fil + V\rfil\inv\ex{\bx}\rfil\right]\,,\label{csm}\\
&V\sm = \left[V\fil\inv + V\rfil\inv\right]\inv.\label{cVs}
\end{align}
Using the definition of the retrofiltered informative mean and information matrix in \erfs{zr}{Lr}, the equations 
can be simplified to 
\begin{align}
\ex{\bx}\sm &= V\sm\left[V\fil\inv\ex{\bx}\fil + {\bf z}\rfil\right]\,,\label{wvsm}\\
V\sm &= \left[V\fil\inv + \Lambda\rfil\right]\inv.\label{wvsV}
\end{align}
We can see that the smoothed state is more accurate than the filtered state through the covariances, where 
it is simple to see that $V\fil \geq V\sm$ in the $N = 1$ case.

\section{LGQ State Estimation}
\label{sec-LGQ}
\subsection{Unconditioned Quantum State}
In the quantum state estimation, we are concerned with estimating a density operator $\rho$ of a quantum 
system as opposed to a PDF $\wp({\blk \bcx})$. For an open quantum system, the evolution of the state $\rho$, 
without observation, is governed by the Lindblad master equation $\hbar\dot\rho = {\cal L} \rho$, with the initial 
condition $\rho(t_0) = \rho_0$, where the Lindbladian superoperator ${\cal L}$ is  
\beq\label{LME}
\quad {\cal L}\bullet = -i[\hat H,\bullet] +{\cal D}[\hat{\bf c}]\bullet\,.
\eeq
Here the Hamiltonian $\hat H$ describes the unitary dynamics of the system and 
$\hat{\bf c} \equiv (\hat{c}_1,\hat{c}_2,...,\hat{c}_M)\!\tp$ is the 
vector of Lindblad operators describing the interacting channels between the system and the environment. {\blk It 
will also be useful to define the row vector form of $\hat{\bf c}$, which we denote by 
$\hat{\bf c}\tp \!= (\hat{c}_1,\hat{c}_2,...,\hat{c}_M)$, where the reader should notice that the transpose does not act 
on the operators within the vector.  Furthermore, the conjugate transpose is defined as the row vector 
$\hat{\bf c}\dg = (\hat{c}_1\dg,\hat{c}_2\dg,...,\hat{c}_M\dg)$. Thus to obtain a column vector form for 
$\hat{\bf c}\dg$, we need to take the transpose. To denote this we will adopt the double dagger notation of 
Ref.~\cite{ChiWis11}, i.e., $\hat{\bf c}\ddg = (\hat{c}_1\dg,\hat{c}_2\dg,...,\hat{c}_M\dg)\!\tp$. 
We can now express the} nonunitary part of \erf{LME} as {\blk
\beq
{\cal D}[\hat{\bf c}]\bullet = \hat{\bf c}\tp\!\!\bullet\hat{\bf c}\ddg - \{\hat{\bf c}\dg\hat{\bf c}/2,\bullet\}\,,
\eeq 
where $\{A,B\} = AB + BA$ is the anticommutator.} Without monitoring the 
environment to gain information about the quantum system, a solution to \erf{LME} is the most accurate estimate 
of the system's quantum state.

We now assume that we 
can describe the quantum system by $N$ bosonic modes. From this we define a 
vector of $2N$ operators $\hbx = (\hat{q}_1,\hat{p}_1,...,\hat{q}_N,\hat{p}_N)\!\tp$, where $\hat{q}_k$ and 
$\hat{p}_k$ are the canonical position and conjugate momentum operators, respectively, describing the 
$k$th bosonic mode and satisfying the commutation relation $[\hat q_k,\hat p_l] = i\hbar\delta_{kl}$.
Furthermore, we assume that the system's Hamiltonian is quadratic and the vector of Lindblad 
operators is linear in $\hbx$, i.e.,~$\hat{H} = \hbx\tp G\hbx/2$ and $\hat{\bf c}= (I_N,iI_N)\bar{C}\hbx$, 
where $G$ and $\bar{C}$ are constant real matrices and $I_n$ denotes an $n\times n$ identity matrix. These 
assumptions ensure that a state initially prepared in a Gaussian state will remain Gaussian throughout the 
evolution. By a Gaussian state we mean one whose Wigner function is Gaussian, 
$W(\bcx) = g(\bcx;\ex{\hbx},V)$, with mean $\ex{\hbx}$ and covariance $V$. The mean 
and covariance are defined as $\ex{\hat x_k} = \Tr[\hat x_k\rho]$ 
and $V_{k,l} = \Tr[\{\hat x_k\hat x_l + \hat x_l \hat x_k\}\rho/2] - \ex{\hat x_k}\ex{\hat x_l}$, 
respectively, where $\hat x_k$ is an element of $\hbx$. For any state $\rho$ the covariance matrix will satisfy 
the \SHUR~\cite{WisMil10},
\beq\label{SHUR}
V+i\hbar\Sigma/2 \geq 0\,,
\eeq
where $\Sigma_{kl} = -i[\hat x_k,\hat x_l]/\hbar$ is a real symplectic matrix.

With these assumptions we can calculate the evolution of the unconditioned LGQ state via its mean and 
covariance, 
\begin{align}
&{\rm d}{\ex \hbx} = A{\ex\hbx} {\rm d}t ,\label{LLE}\\
&\frac{{\rm d}V}{{\rm d}t} = AV +V A\tp \!+ D\,,
\label{UncondV}
\end{align}
with the initial conditions for the mean and covariance $\ex{\hbx}(t_0) = \ex{\hbx}_0$ and $V(t_0) = V_0$, 
respectively. Here the drift and diffusion matrices are \cite{WisMil10}
\beq
A = \Sigma(G+\bar{C}\tp \!S\bar{C})\,, \qquad D = \hbar\Sigma\bar{C}\tp\bar{C}\Sigma\tp\,,
\eeq
respectively, with $S = \left[\begin{smallmatrix}
0&I_N\\
-I_N&0
\end{smallmatrix}\right]$ being another symplectic matrix.

\subsection{Filtered Quantum State} \label{Mintro}
In order to obtain a better estimate of the system's state than the unconditioned state, 
we need to gain more information about 
the system by measuring the environment. In this work we focus on diffusive-type 
unravelings of the master equation as opposed to a jump unraveling, as the former 
preserves Gaussian states. The 
corresponding stochastic master equation, sometimes referred to as a quantum filtering equation 
\cite{Belav87,Bel92} for reasons that will become apparent, in the $M$ representation \cite{ChiWis11} is
\beq\label{SME}
\hbar\dd \rho\fil = {\cal L}\rho\fil\dd t + \sqrt{\hbar}\dd\bw\fil\tp{\cal H}[M\dg \hat{\bf c}]\rho\fil\,.
\eeq
Here, 
${\cal H}[\hat{\bf a}]\bullet =\hat{\bf a}\bullet+\bullet\hat{\bf a}\ddg-\Tr[\bullet(\hat{\bf a}+\hat{\bf a}\ddg)]\bullet$, 
and the initial condition is $\rho\fil(t_0) = \rho_0$. 
We have also implicitly introduced a vector of measurement currents 
$\by\dd t = \langle M\dg\hat{\bf c} + M\tp\hat{\bf c}\ddg\rangle\fil\dd t + \dd\bw\fil$ where 
$\ex{\bullet}\fil := \Tr[\bullet \rho\fil]$ through the vector of innovations $\dd\bw\fil$, which satisfies similar 
conditions to \erf{WeinCond}. 

To ensure that evolution under \erf{SME} does not result in an 
invalid quantum state, it is necessary and sufficient \cite{ChiWis11} for 
$M$ to satisfy $MM\dg = {\rm diag}(\eta_1,\eta_2,...,\eta_M)$, where $\eta_k$ can be interpreted as the 
monitoring efficiency of the channel $\hat{c}_k$.
Note, we can also define an un-normalized filtered state $\tilde\rho\fil$, which explicitly depends on the 
measurement results $\by\dd t$ (instead of the innovation $\dd\bw\fil$), reflecting the observer's 
knowledge of the system. This un-normalized filtered state satisfies the stochastic master equation
\beq\label{USME}
\hbar\dd \tilde\rho\fil = {\cal L}\tilde\rho\fil\dd t + \sqrt{\hbar}\by\tp\widetilde{\cal H}[M\dg \hat{\bf c}]
\tilde\rho\fil\dd t\,,
\eeq
where $\widetilde{\cal H}[\hat{\bf a}]\bullet =\hat{\bf a}\bullet+\bullet\hat{\bf a}\ddg$.

Restricting the discussion to LGQ systems, we can express the vector of measurement current as 
\beq
\by\dd t = C\ex{\hat\bx}\fil\dd t + \dd\bw\fil\,,
\eeq
where $C = 2\sqrt{\hbar\inv}  T\tp\bar{C}$, $T\tp \!= ({\rm Re}[M\tp],{\rm Im}[M\tp])$, and 
${\rm d}\bw\fil \equiv \by{\rm d}t - C\ex{\hbx}\fil{\rm d}t$. From the stochastic 
master equation in \erf{SME}, we can derive the equations for the mean and covariance of the filtered 
state, giving
\begin{align}
&{\rm d}\ex{\hbx}\fil=\, A\ex{\hbx}\fil {\rm d}t + {\cal K}^{+}[V\fil]{\rm d}\bw\fil\,,\label{qfm}\\
&\frac{{\rm d}V\fil}{{\rm d}t} = \, AV\fil +V\fil A\tp \!+ D - {\cal K}^{+} [V\fil] {\cal K}^{+} [V\fil]\tp\,,
\label{qVf}
\end{align}
with initial conditions $\xfil(t_0) = \ex{\hbx}_0$ and $V\fil(t_0) = V_0$.
The optimal Kalman gain matrix, $\K^+[V\fil]$, which we will later refer to as a {\em kick} matrix,
is defined in \erf{Kick}, with the 
measurement back-action $\Gamma = -\sqrt{\hbar}T\tp \!S\bar{C}\Sigma\tp$. Note that these 
equations for the filtered quantum state have exactly the same form as the classical Kalman-Bucy filtering 
equations.

\subsection{Retrofiltered Effect and {\red Smoothed Wigner Distribution}}
The retrofiltered effect gives the probability density of a measurement result occurring at a later time 
given a particular quantum state at the current time: 
\beq
\wp(\fut{\blk\rm O}|\rho) = \Tr[\rho\hat E\rfil]\,, 
\eeq
where $\hat{E}\rfil$ is a function of the future record $\fut{\blk\rm O}$. 
The effect $\hat{E}\rfil$ can be computed backward in time from a final uninformative effect 
$\hat E\rfil(T) \propto \hat I$. The 
stochastic equation for the (unnormalized) retrofiltered effect $\hat{E}\rfil$ is obtained 
by taking the adjoint of \erf{USME}, giving
\beq\label{USEE}
-\hbar\dd\hat{E}\rfil = {\cal L}\dg \hat{E}\rfil \dd t + \sqrt{\hbar}\by\widetilde{{\cal H}}[M\tp\hat{\bf c}\ddg]
\hat{E}\rfil\dd t\,,
\eeq
where ${\cal L}\dg$ is the adjoint of the Lindbladian superoperator. Note that \erf{USEE} is not 
trace-preserving and evolves backward in time. Following a similar logic to that presented in the classical 
case, we will normalize the retrofiltered effect, as ultimately we are interested in a smoothed state which will 
require normalization regardless. In doing so, we obtain a normalized retrofiltered effect $\hat{E}\rfil'$
\cite{ZhaMol17},
\beq
\begin{split}
-\hbar\dd \hat{E}'\rfil = {\cal L}\dg \hat{E}'\rfil \dd t - \ex{\hat\kappa}&\rfil\hat{E}'\rfil\dd t \\+
& \sqrt{\hbar}\dd\bw\rfil{\cal H}[M\tp\hat{\bf c}\ddg]
\hat E\rfil'\,,
\end{split}
\eeq
where $\dd\bw\rfil = \by\dd t - \ex{M\dg\hat{\bf c} + M\tp\hat{\bf c}\ddg}\rfil\dd t$ with
$\ex{\bullet}\rfil := \Tr[\bullet \hat{E}'\rfil]$ and $\hat\kappa = \hat{\bf c}\tp\hat{\bf c}\ddg - \hat{\bf c}\dg\hat{\bf c}$.
 
Considering an LGQ system, the Wigner function for the normalized retrofiltered effect is a normalized 
Gaussian, i.e.,~$W\rfil(\bcx) = g(\bcx;\ex{\hbx}\rfil,V\rfil)$. Consequently, we can obtain, in a similar way to the 
filtered case in \erfs{qfm}{qVf}, the equations for the retrofiltered mean and covariance,
\begin{align}
&- {\rm d}\ex{\hbx}\rfil =  -A\ex{\hbx}\rfil {\rm d}t + {\cal K}^{-} [V\rfil]{\rm d}\bw\rfil,\label{qrm}\\
-& \frac{{\rm d}V\rfil}{{\rm d}t} = -AV\rfil - V\rfil A\tp \!+ D - {\cal K}^{-} [V\rfil] {\cal K}^{-} [V\rfil]\tp.\label{qVr}
\end{align}
These equations completely describe the effect, with the final condition $V\rfil(T) = \infty$. Once again, there 
is no sensible final condition for the retrofiltered mean due to the infinite covariance. Following the 
same procedure presented in the classical case, we obtain \erfs{zr}{Lr}, where in the quantum case 
${\bf z}\rfil := \Lambda\rfil\ex{\hbx}\rfil$.

Following the classical equations, one might think that we could obtain a Gaussian smoothed quantum state 
$W\swd(\bcx) = g(\bcx;\ex{\hbx}\swd,V\swd)$, with mean $\ex{\hbx}\swd$ and covariance $V\swd$ given by
\begin{align}
\ex{\hbx}\swd &= V\swd\left[V\fil\inv\ex{\hbx}\fil + V\rfil\inv\ex{\hbx}\rfil\right]\,,\label{wvsm}\\
V\swd &= \left[V\fil\inv + V\rfil\inv\right]\inv.\label{wvsV}
\end{align}
While this construction might seem valid, {\blk  we 
will show using an example in Sec.~\ref{sec-Traj} that the \red SWD \blk covariance does not always satisfy the 
\SHUR~\erf{SHUR}, as it would if it were a valid quantum state.} Thus we turn to 
quantum state smoothing theory instead.


\subsection{LGQ State Smoothing}
For the quantum state smoothing theory \cite{GueWis15}, we consider an open quantum system coupled 
to two baths. In principle, each of these baths can comprise any number of physically distinct baths, but for 
simplicity we will consider them collectively. An 
observer, Alice, monitors one of the baths and is able to construct a measurement record $\blk\rm O$, which 
we will refer to as the `observed' record. A (perhaps hypothetical) secondary observer, Bob, monitors the 
remaining bath and constructs his own measurement record $\blk\rm U$ that is unobserved by Alice, which we 
will call the `unobserved' record. See Fig.~\ref{Fig-QSS}. 
Now Bob, assumed to have access to both the observed and the unobserved record, can estimate the 
quantum state conditioned on both $\past{\blk\rm O}$ and $\past{\blk\rm U}$.
That is, he obtains a state with maximal information about the 
quantum system, which can be regarded as the {\em true} state 
$\rho\god:=\rho_{\past{\blk\rm O}\past{\blk\rm U}}$. However, since 
Alice does not have access to $\past{\blk\rm U}$, she can only obtain an estimate of the true state based on 
her observed measurement record. In this case she can construct a conditioned state with the form
\beq \label{QSS}
\rho\c = \sum_{\past{\blk\rm U}} \wp\c(\past{\blk\rm U}) \rho\god\,,
\eeq
where the conditioning `${\rm C}$' depends on the amount of the observed measurement record used in 
the estimation. 
If Alice wishes to obtain a filtered state, i.e., ${\rm C} \equiv \rm{F}$, the conditioned probability distribution 
for the unobserved record becomes $\wp\fil(\past{\blk\rm U}) = \wp(\past{\blk\rm U}|\past{\blk\rm O})$. To 
obtain a smoothed state, i.e., ${\rm C} \equiv \rm{S}$, the conditional probability becomes 
$\wp\sm(\past{\blk\rm U}) = \wp(\past{\blk\rm U}|\both{\blk\rm O})$.

For LGQ state smoothing \cite{LCW19}, the true state of the system is represented by 
a Gaussian Wigner function $W\god(\bcx) = g(\bcx;\ex{\hbx}\god,V\god)$. We introduce an 
unobserved measurement current 
$\by\un\dd t= C\un \ex{\hbx}\god\dd t + \dd\bw\un$ to account for Bob's monitoring of the environment, in 
addition to Alice's observed measurement current $\by\ob\dd t = C\ob\ex{\hbx}\god\dd t + \dd\bw\ob$, 
where $\dd\bw\un$ and $\dd\bw\ob$ are the unobserved and observed innovations, respectively. The true 
state of the system can be obtained by conditioning the estimate on both Alice's and Bob's past 
measurement records, giving 
\begin{align}
&{\rm d}\xgod = A\xgod{\rm d}t + {\cal K}^{+}\ob[V\god]{\rm d}\bw\ob + {\cal K}^{+}\un[V\god]{\rm d}
\bw\un\,,\label{truest}\\
&\frac{{\rm d}V\god}{{\rm d}t} =  A V\god + V\god A\tp \!+ D \nn\\
&\qquad\qquad- {\cal K}^{+}\ob[V\god]{\cal K}^{+} \ob[V\god]\tp - {\cal K}^{+}\un[V\god]
{\cal K}^{+}\un[V\god]\tp\,, 
\label{truvar}
\end{align}
where ${\cal K}^\pm_{\rm r} [V] = VC\tp_{\rm r} \!+ \Gamma\tp_{\rm r}$ for r $\in \{$o,u$\}$ and the initial 
conditions are $\xgod(t_0) = \ex{\hbx}_0$ and $V\god(t_0) = V_0$. This follows 
trivially by extending \erfs{qfm}{qVf} to two measurement records.

Since we are restricting to Gaussian states, the true state depends on $\past{\blk\rm U}$ only via the 
mean in \erf{truest}. This means that we can replace the (symbolic) summation in \erf{QSS} by an integral 
over the true mean, so that the smoothed state (${\rm C=S}$) is given by
\beq\label{conv1}
\rho\sm = \int \wp\sm(\xgod) \rho\god(\ex{\hat\bx}\god) {\rm d}\xgod\,,
\eeq
where the PDF $\wp\sm(\xgod)$ is for the true mean conditioned on the past-future observed record.

We can replace the 
smoothed state and the true state by their Wigner functions, the latter of which is replaced by a Gaussian 
$g(\bcx;\halo\bx,V\god)$. Here we have defined a haloed variable $\halo\bx = \ex{\hbx}\god$ for notational 
simplicity\footnote{\blk We use this halo notation because these haloed variables are effectively a mediary 
between an estimate known only to an omniscient observer (i.e., the true state) and estimates available to partially 
ignorant observers (e.g. the smoothed state).}. To obtain the smoothed state in \erf{conv1}, we convolve the true 
state with the conditional PDF 
(which is a classically smoothed LG distribution) $\wp\sm(\halo\bx) = g(\halo\bx;\ex{\halo\bx}\sm,\hV\sm)$, 
where $\ex{\halo\bx}\sm$ and $\hV\sm$ will be determined later. Since both functions in the convolution are 
Gaussian, the resulting smoothed state is also Gaussian. Consequently, we can rewrite \erf{conv1} as 
\beq
g(\bcx;\ex{\hat\bx}\sm,V\sm) = \int g(\halo\bx;\ex{\halo\bx}\sm,\halo V\sm) g(\bcx;\halo\bx,V\god) {\rm d}
\halo\bx\,.
\eeq
From the properties of a Gaussian convolution, we find that $\ex{\hbx}\sm = \ex{\halo\bx}\sm$ and 
$V\sm = \hV\sm + V\god$. 

All that remains is to determine the haloed mean and covariance of the 
smoothed Gaussian PDF $\wp\sm(\xgod)$. By rewriting the equation for the true mean, \erf{truest}, as 
\beq\label{hLLE}
\dd\halo\bx = A\halo\bx\dd t + \halo E\dd\halo\bv\p\,,
\eeq
where $\halo{E}\dd\halo\bv\p = \K^+\ob[V\god]\dd\bw\ob + \K^+\un[V\god]\dd\bw\un$, we see that the 
system evolves according to a classical linear Langevin equation of the form in \erf{LLE}. Furthermore, the 
observed measurement record $\by\ob = C\ob\halo\bx + \dd\bw\ob$ is linear in $\halo\bx$ and we can 
define a new cross-correlation 
$\halo\Gamma\tp = \K^+\ob[V\god]$. Since the PDF satisfies the requirements for classical LG state 
estimation, we can use \erfs{csm}{cVs} and obtain the haloed smoothed mean 
and covariance, given by 
\begin{align}
\ex{\halo\bx}\sm &= \hV\sm\left[\hV\fil\inv\ex{\halo\bx}\fil + \hV\rfil\inv\ex{\halo\bx}\rfil\right]\,,\label{hxs}\\
\hV\sm &= \left[\hV\fil\inv + \hV\rfil\inv\right]\inv\,.\label{hVs}
\end{align}
We can obtain the haloed filtered mean and covariance, $\ex{\halo\bx}\fil$ and $\hV\fil$, and haloed 
retrofiltered mean and covariance, $\ex{\halo\bx}\rfil$ and $\hV\rfil$, by conditioning $\halo\bx$ on the past 
observed and future observed measurement records, respectively. 

By conditioning \erf{hLLE} on only the past observed measurement record, we 
obtain the haloed filtered variables
\begin{align}
{\rm d}\ex{\halo\bx}\fil = &\,\, A\ex{\halo\bx}\fil{\rm d}t + {\cal K}^{+}\ob[\hV\fil+V\god] 
{\rm d}\halo\bw\fil\,,\\
\ddt{\hV\fil} = & \,\, A\hV\fil + \hV\fil A\tp \!+ \halo D \nn\\
&- {\cal K}^{+}\ob[\hV\fil +V\god]{\cal K}^{+}
\ob[\hV\fil +V\god]\tp\,,\label{hVfil}
\end{align}
where $\halo D = \K^+\ob[V\god]\K^+\ob[V\god]\tp \!+ \K^+\un[V\god]\K^+\un[V\god]\tp$ and 
${\rm d}\halo\bw\fil = \by\ob{\rm d}t - C\ob\ex{\halo\bx}\fil {\rm d}t$. From \erf{qVf} and (\ref{hVfil}), 
it can easily be shown that $\halo V\fil = V\fil - V\god$, and using this relationship we can show that 
$\ex{\hbx}\fil = \ex{\halo\bx}\fil$. Similarly, the haloed retrofiltered variables are given by
\begin{align}
-{\rm d}\ex{\halo\bx}\rfil = &-A\ex{\halo\bx}\rfil{\rm d}t + {\cal K}^{-}\ob[\hV\rfil - V\god]{\rm d}
\halo\bw\rfil\,,\\
-\ddt{\hV\rfil} = & -A\hV\rfil - \hV\rfil A\tp \!+ \halo D \nn\\
& - {\cal K}^{-}\ob[\hV\rfil -V\god]{\cal K}^{-}\ob[\hV\rfil - V\god]\tp\,,\label{hVR}
\end{align}
where $\dd\halo\bw\rfil = \by\ob\dd t - C\ob\ex{\halo\bx}\rfil$. It can be shown, using \erf{qVr}, that 
$\halo V\rfil = V\rfil + V\god$, and from this we can also show 
that $\ex{\hbx}\rfil = \ex{\halo\bx}\rfil$. 
Finally, using \erfs{hxs}{hVs}, we can compute the mean and covariance of the smoothed 
quantum state:
\begin{align}
\xsm = (V\sm - V\god) [(V\fil &- V\god)\inv\ex{\hbx}\fil \nn\\
&+ (V\rfil+V\god)\inv \ex{\hbx}\rfil]\,,\label{estsm}\\
V\sm = \big[(V\fil - V\god )\inv &  + (V\rfil + V\god)\inv\big]\inv + V\god\,. \label{varsm}
\end{align}
Interestingly, we notice that the equations for the smoothed quantum state are similar to the equations for 
the \red SWD \blk state in \erfs{wvsm}{wvsV}. In fact they are identical if we allow for $V\god\to 0$, which is 
equivalent to a classical limit where we set $\hbar\to 0$ in \erf{SHUR}. Unsurprisingly, we can 
see that the LGQ smoothed covariance places less emphasis on the retrofiltered covariance than the \red SWD \blk 
covariance. This can be seen from \erf{varsm} where $(V\rfil + V\god)\inv$ is smaller than 
$V\rfil\inv$. The reason this is unsurprising is because combining the filtered covariance with the 
retrofiltered covariance resulted in the \red SWD \blk covariance violating the \SHUR, which is avoided when 
combining with the smaller $(V\rfil + V\god)\inv$.

As was the case with the retrofiltered mean, the haloed retrofiltered mean $\ex{\halo\bx}\rfil$ does not 
have a well defined final condition due to the haloed retrofiltered covariance being infinite at the final time. 
However, we can solve this problem in 
the same way as we did for the retrofiltered mean and covariance by defining the haloed retrofiltered 
informative mean $\halo{\bf z}\rfil = \hL\rfil\ex{\halo\bx}\rfil$ and corresponding information matrix 
$\hL\rfil = \hV\rfil\inv$. 
Using \erf{V2L}, we obtain
\begin{align}
-\dd\halo{\bf z}\rfil =&\, (\bar{A} - \bar{D}\hL\rfil)\!\tp \halo{\bf z}\rfil\dd t \nn\\&+ (C\ob\tp \!- \hL\rfil V\god 
C\ob\tp \!- \hL\rfil \Gamma\tp) \by\ob\dd t\,,\\
-\frac{\dd \hL\rfil}{\dd t} =&\, \hL\rfil\bar{A} + \bar{A}\tp\hL\rfil - \hL\rfil\bar{D}\hL\rfil + C\ob\tp C\ob\,,
\end{align}
where $\bar{A} = A - \Gamma\ob\tp C\ob - V\god C\ob\tp C\ob$ and 
$\bar{D} = \K^{+}\un[V\god]\K^{+}\un[V\god]\tp$. The final conditions become $\halo{\bf z}\rfil(T) = 0$ and 
$\hL\rfil(T) = 0$. With these definitions, we can further simplify the LGQ smoothing equations, 
\erfs{estsm}{varsm}, to 
\begin{align}
\xsm = (V\sm - V\god) [(V\fil &- V\god)\inv\ex{\hbx}\fil + \halo{\bf z}\rfil]\,,\label{estsm}\\
V\sm = \big[(V\fil - V\god )\inv &  + \hL\rfil\big]\inv + V\god\,. \label{varsm}
\end{align}

\section{Physical LGQ Systems}
\label{sec-PS}
For the remainder of this paper, we will consider two examples of LGQ systems: an on-threshold optical 
parametric oscillator and a {\blk noisy linear attenuator}.
In both examples, Alice and Bob perform homodyne measurements on the environment, where we 
use measurement efficiencies to quantify the fraction of the environment that they can observe. 

\subsection{On-Threshold Optical Parametric Oscillator}\label{Sec-OPO}
The first system we consider is an optical parametric oscillator (OPO) with one output 
channel (loss, at rate unity). This is described by the master equation 
\beq
\hbar \dot\rho = i\chi[(\hat q\hat p + \hat p \hat q)/2,\rho] + \gamma{\cal D}[(\hat q +i\hat p)] \rho\,,
\eeq
where the number of modes is $N = 1$ and $\hbx = (\hat{q},\hat{p})\!\tp$. We will consider the on-threshold 
parameter regime, when $\chi = \gamma$ and for simplicity we measure time in units of $\chi\inv$. The first 
term is generated by the squeezing Hamiltonian $\hat{H} = (\hat q\hat p + \hat p \hat q)/2$ 
and the second term is the Lindblad term with $\hat{\bf c} = \hat q +i\hat p$ describing photon loss. From 
these we find that 
\beq
G = \left(\begin{array}{cc}
0&1\\
1&0
\end{array}\right)\,, \qquad \bar{C} = I_2\,,
\eeq 
by remembering that $\hat{H} = \hbx\tp G\hbx/2$ and $\hat{\bf c} = (I_2,iI_2)\bar{C}\hbx$. We then find the 
drift and diffusion matrices $A = {\rm diag}(0,-2)$ and $D = \hbar I_2$.

Let us assume that the output (loss) channel is monitored by Alice and Bob using homodyne 
measurements with homodyne phases $\theta\ob$ and $\theta\un$ and measurement efficiencies $\eta\ob$ 
and $\eta\un$, respectively. The resulting measurement current for this type of measurement is 
\beq
\by_{\rm r}\dd t = \sqrt{\eta_{\rm r}}\ex{e^{-i\theta_{\rm r}} a + e^{i\theta_{\rm r}}a\dg}\god + \dd \bw_{\rm r}
\eeq
for ${\rm r} \in \{{\rm o,u}\}$ and the annihilation operator $a = (\hat{q} + i\hat{p})/\sqrt{2}$.
As a result, we can define $M_{\rm r} = \sqrt{\eta_{\rm r}}e^{i\theta_{\rm r}}$, where $M$ is the 
{\blk unraveling} 
matrix introduced in \erf{SME}. Thus, Alice's measurement and back-action matrices are 
$C\ob = 2\sqrt{\eta\ob/\hbar} (\cos\theta\ob,\sin\theta\ob)$ and 
$\Gamma\ob = -\hbar C\ob/2$, respectively. Similarly, Bob's unobserved measurement and back-action 
matrices are $C\un =  2\sqrt{\eta\un/\hbar} (\cos\theta\un, \sin\theta\un)$ and 
$\Gamma\un = -\hbar C\un/2$, respectively. 

\subsection{Noisy Linear Attenuator}\label{ssec-LA}
The second system we consider is a single-mode ($N=1$) {\blk noisy linear attenuator}, 
described by the master equation
\beq
\hbar \dot\rho = \gamma\dwn{\cal D}[\hat q + i\hat p]\rho + \gamma\up{\cal D}[\hat q - i\hat p]\rho\,,
\eeq
where $\gamma\dwn$ and $\gamma\up$ are the rate of photon loss and gain, respectively.
The fact that this system acts as an attenuator can be seen in how the annihilation operator changes, on 
average, over time,
\beq
\ex{\dot{\hat{a}}} = (\gamma\up - \gamma\dwn)\ex{\hat a}\,,
\eeq
where for the system to be classed as an attenuator and not an 
amplifier, we consider the case when $\gamma\dwn>\gamma\up$.

Since there are no Hamiltonian dynamics for this system, i.e.,~$G = 0$, we only need to concern 
ourselves with the vector of Lindblad operators,
\beq
\hat{\bf c} = \left[\sqrt{\gamma\dwn}(\hat q + i\hat p),\sqrt{\gamma\up}(\hat q - i\hat p)\right]\tp\,.
\eeq 
Note, this vector of Lindblad operators is not to be confused with a commutator.
From this, we calculate 
\beq
\bar{C} = \left(\begin{array} {cccc}
\sqrt{\gamma\dwn}&\sqrt{\gamma\up}&0&0\\
0&0&\sqrt{\gamma\dwn}&-\sqrt{\gamma\up}\\
\end{array}\right)\tp\
\eeq
and arrive at $A = (\gamma\up - \gamma\dwn)I_2$ and $D = \hbar(\gamma\up + \gamma\dwn)I_2$. 

In this case, since we are considering homodyne measurements on both channels, we can take 
$M_{\rm r} = {\rm diag}(\sqrt{\eta_{\downarrow,{\rm r}}}e^{i\theta_{\downarrow,{\rm r}}}, 
\sqrt{\eta_{\uparrow,{\rm r}}}e^{i\theta_{\uparrow,{\rm r}}})$ for ${\rm r} \in \{{\rm o},{\rm u}\}$. Here we 
have introduced the measurement efficiencies $\eta_{\downarrow,{\rm r}}$ and $\eta_{\uparrow,{\rm r}}$ 
for the attenuation and the amplification channels, respectively, to indicate the fraction of each output that 
is measured by Alice (o) and Bob (u), with the homodyne phases $\theta_{\downarrow,{\rm r}}$ and 
$\theta_{\uparrow,{\rm r}}$. The measurement and back-action matrices, for either Alice or Bob, are given 
by
\beq
C_{\rm r} = \frac{2}{\sqrt{\hbar}}\left(\begin{array}{cc}
\sqrt{\eta_{\downarrow,{\rm r}}\gamma\dwn}\cos\theta_{\downarrow,{\rm r}} & 
\sqrt{\eta_{\downarrow,{\rm r}}\gamma\dwn}\sin\theta_{\downarrow,{\rm r}}\\
\sqrt{\eta_{\uparrow,{\rm r}}\gamma\up}\cos\theta_{\uparrow,{\rm r}} & 
-\sqrt{\eta_{\uparrow,{\rm r}}\gamma\up}\sin\theta_{\downarrow,{\rm r}}\\
\end{array}\right)
\eeq
and
\beq
\Gamma_{\rm r} = \sqrt{\hbar}\left(\begin{array}{cc}
-\sqrt{\eta_{\downarrow,{\rm r}}\gamma\dwn}\cos\theta_{\downarrow,{\rm r}} & 
-\sqrt{\eta_{\downarrow,{\rm r}}\gamma\dwn}\sin\theta_{\downarrow,{\rm r}}\\
\sqrt{\eta_{\uparrow,{\rm r}}\gamma\up}\cos\theta_{\uparrow,{\rm r}} & 
-\sqrt{\eta_{\downarrow,{\rm r}}\gamma\up}\sin\theta_{\downarrow,{\rm r}}\\
\end{array}
\right)\,,
\eeq
respectively.

For this system there are many scenarios we could consider for Alice and Bob. For example, Alice and 
Bob could each perfectly monitor one of the channels, or they could both monitor the same output channel 
with some fractions. However, for simplicity, we will only consider the case where Alice perfectly measures 
the attenuation channel, i.e.,~$\eta_{\downarrow,{\rm o}} = 1$ and $\eta_{\uparrow,{\rm o}}=0$, with a 
homodyne phase $\theta_{\downarrow,{\rm o}} = \theta\ob$, and Bob perfectly measures the amplification 
channel, i.e.,~$\eta_{\downarrow,{\rm u}} = 0$ and $\eta_{\uparrow,{\rm u}}=1$, with a homodyne phase 
$\theta_{\uparrow,{\rm u}} = \theta\un$. 

\section{Example Trajectories}
\label{sec-Traj}
\begin{figure*}
\begin{minipage}{.5\textwidth}
\includegraphics[scale = 0.32]{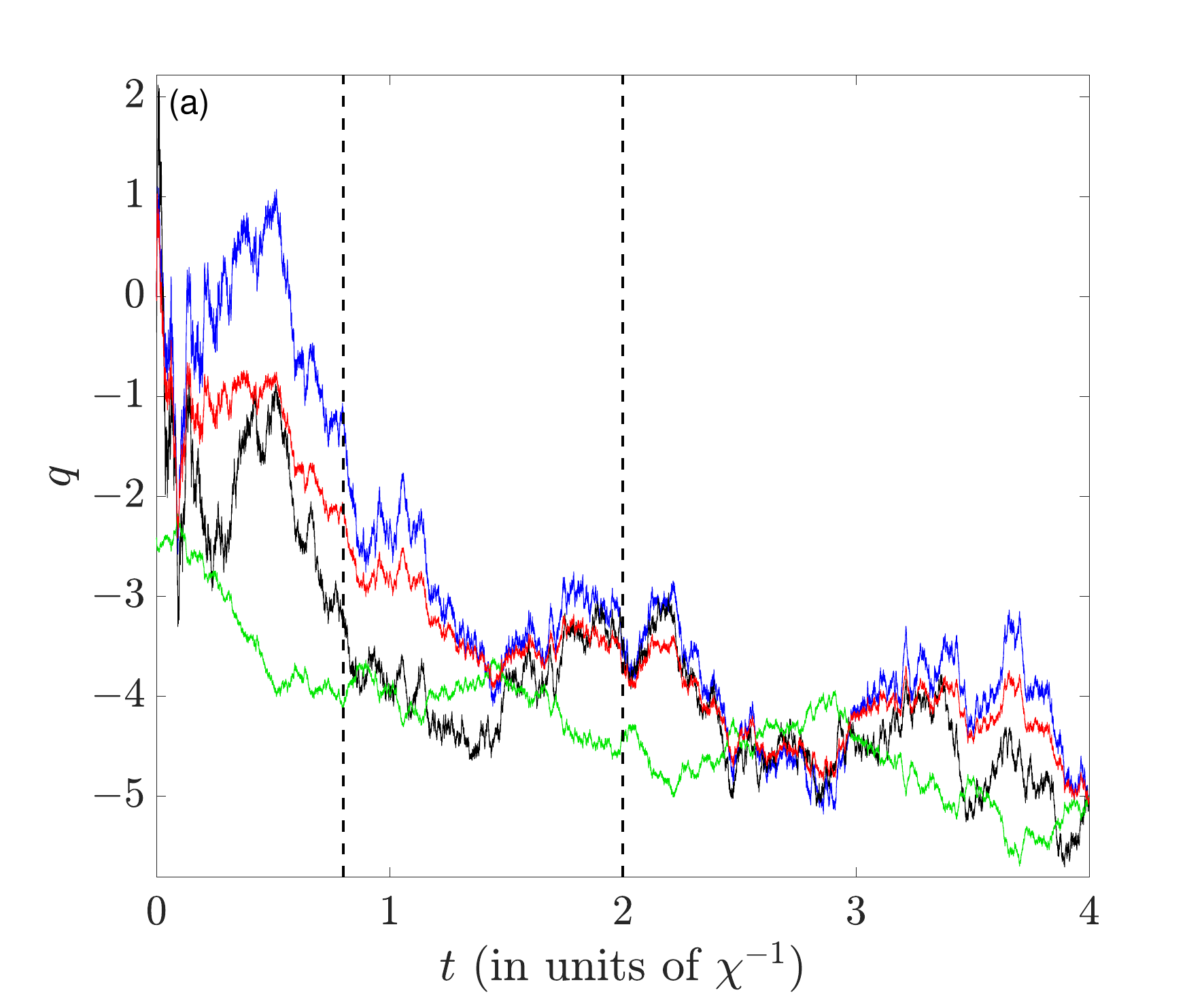}
\includegraphics[scale = 0.32]{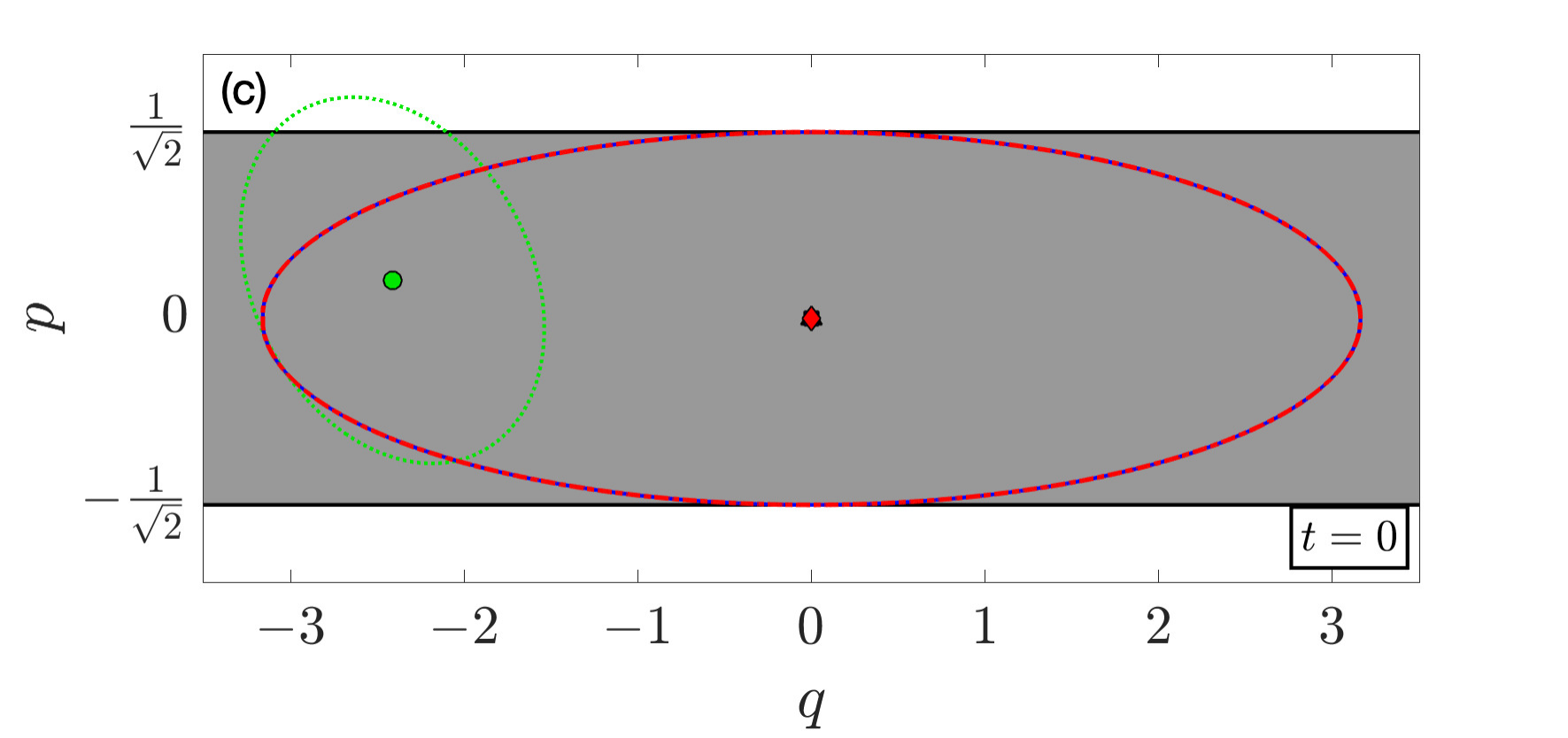}
\includegraphics[scale = 0.32]{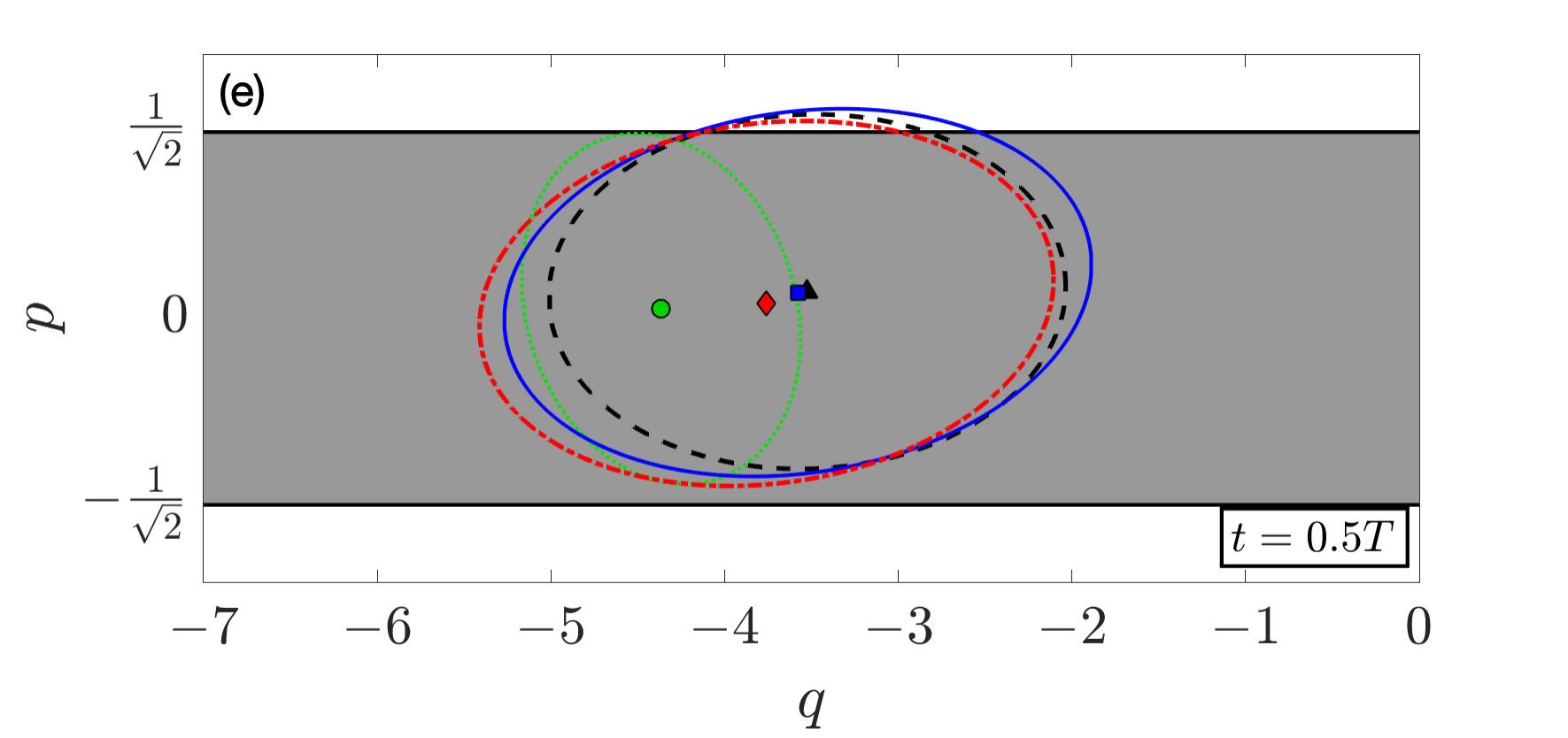}
\end{minipage}%
\begin{minipage}{.5\textwidth}
\includegraphics[scale = 0.32167895]{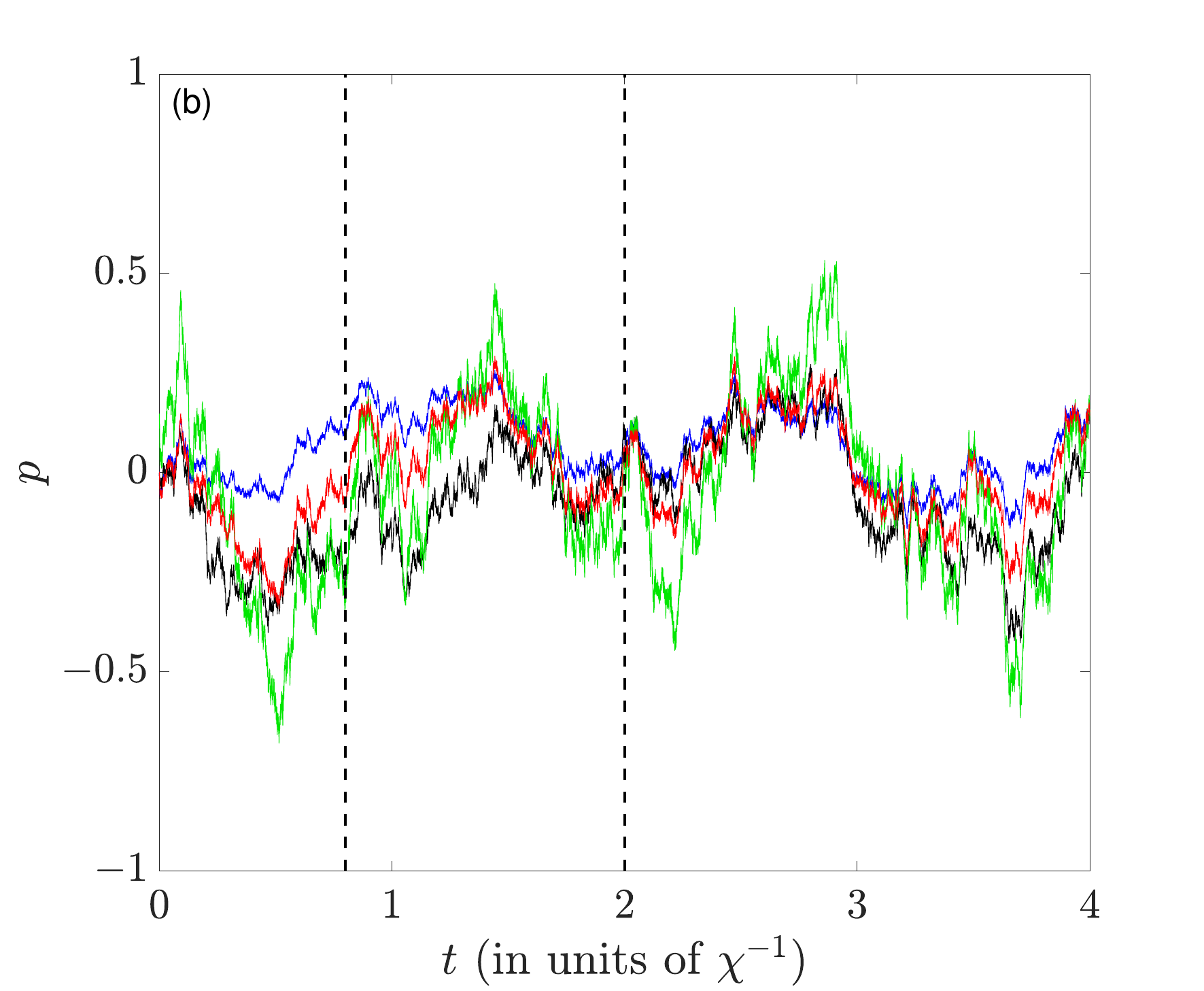}
\includegraphics[scale = 0.32]{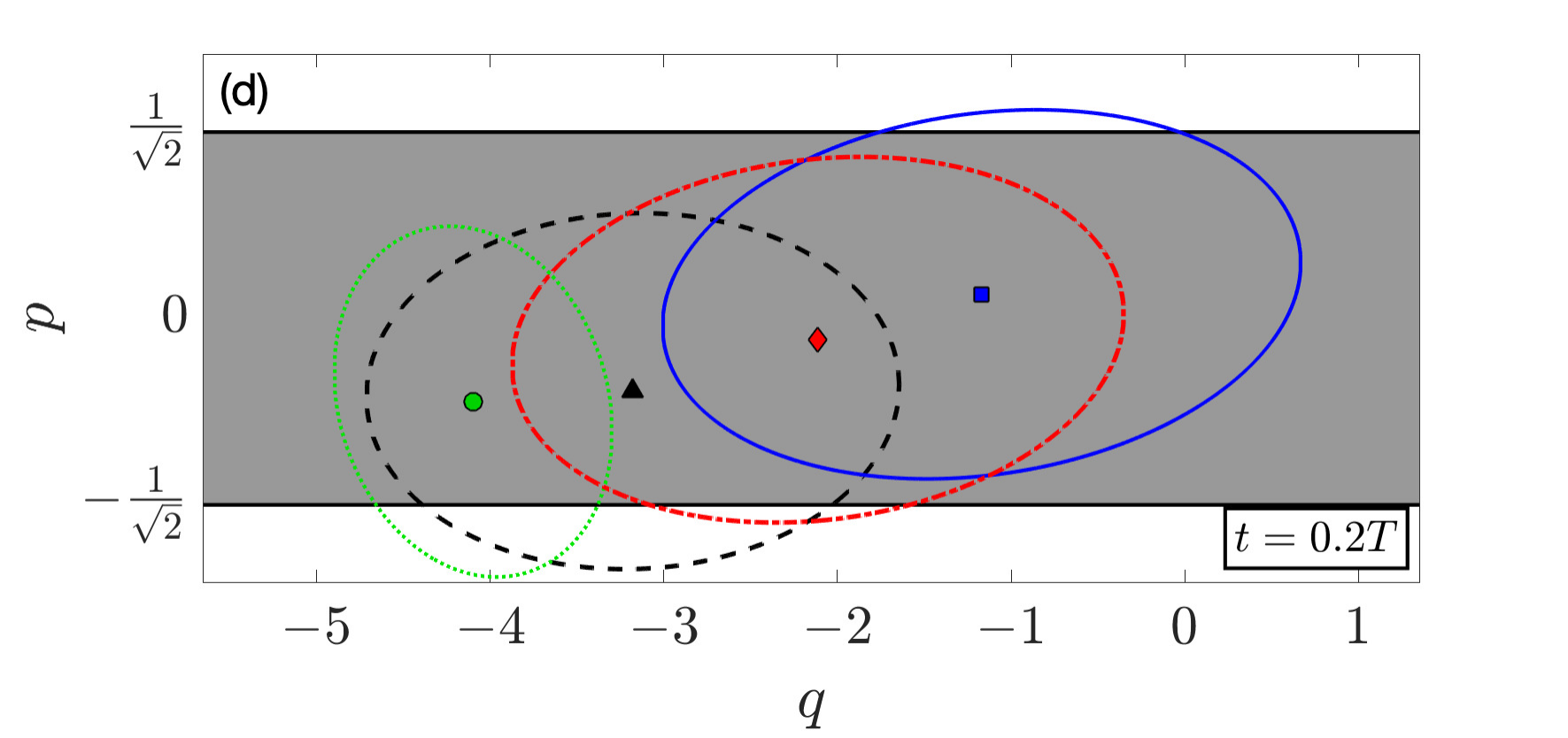}
\includegraphics[scale = 0.32]{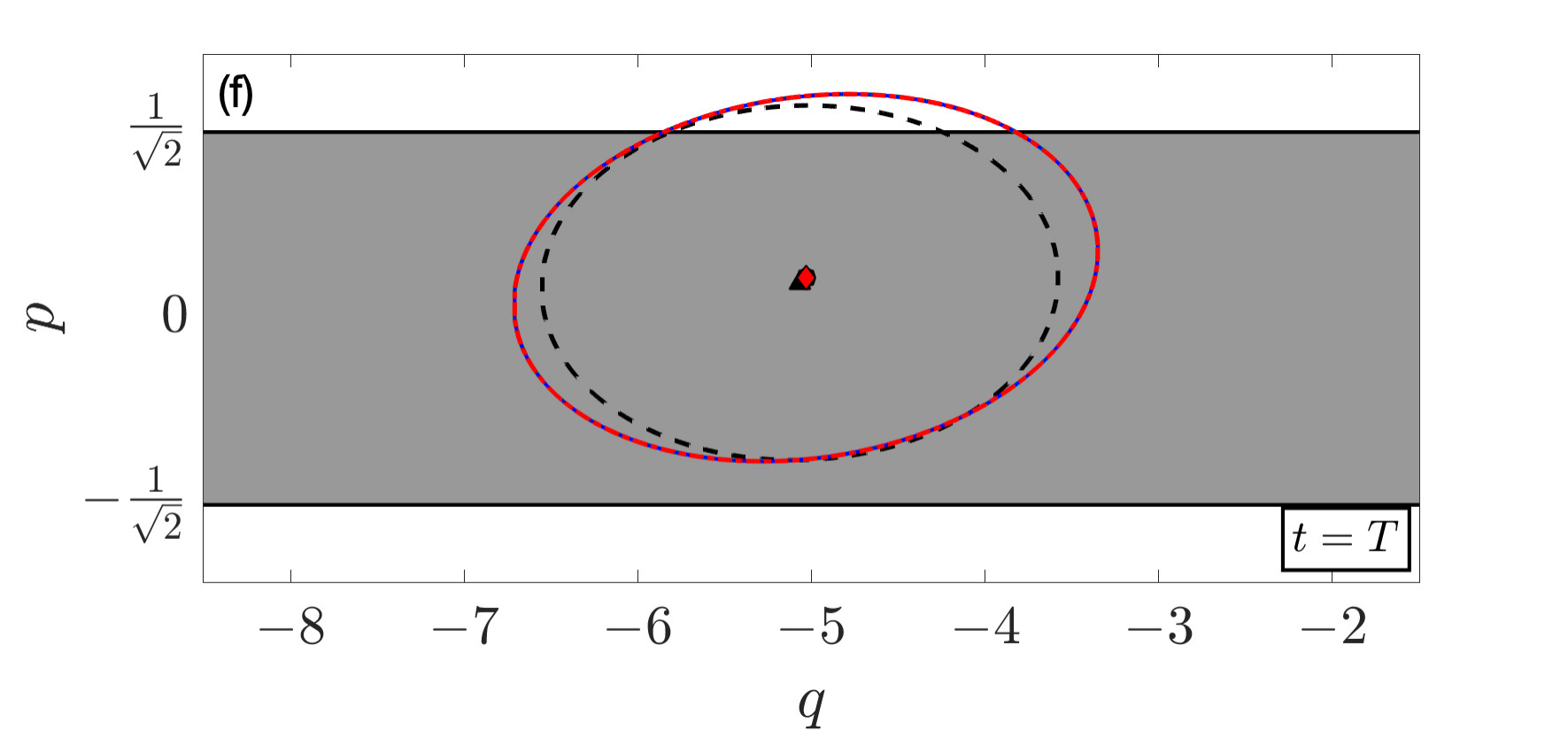}
\end{minipage}
\caption{A sample realization of the OPO system's state trajectory, where $\eta\ob = \eta\un = 0.5$, 
$\theta\ob = \pi/4$, and $\theta\un = -\pi/8$, where the time $t$ is in units of $\chi\inv$ and the total run 
time $T = 4$. We have set $\hbar = 2$ for this simulation.
The evolutions in the $q$ and $p$ quadratures, in panels (a) and (b), 
respectively, clearly show that the smoothed mean (red) outperforms the filtered mean (blue) in terms of 
estimating the true mean (black). The \red SWD \blk mean (green), on the other hand, does a terrible job of 
estimating the true mean, as expected. The disparity between the \red SWD \blk state and the remaining states 
can clearly be seen in the phase-space diagrams, plotted at four snapshots in time in the panels (c)--(f). 
In (c), the filtered, 
smoothed, and true states all begin at the same point, with the same covariance {\red (where the ellipse 
indicates the $e^{-1/2}$ contour of the Wigner function corresponding to one standard deviation along its principle axes and the orientation of the ellipse captures the diagonal basis of the covariance)}. However, the mean of the \red SWD \blk state (green dot) is 
largely displaced from the rest and its 
covariance is significantly smaller. As time progresses, the filtered, smoothed, and true states begin to 
separate and the covariances decrease, where the smoothed covariance sits somewhere between the 
filtered and true covariance. At the final time $T$, only the true state is displaced from the remaining states, 
which are all the same, as there is no future record left.}
\label{Fig-Traj}
\end{figure*}

In this section we will compare the filtered, \red SWD \blk, and smoothed quantum states in order to see the 
differences between these estimated states and how well they estimate the true state.
We will only consider the OPO system in this section, since the results are similar for the {\blk noisy linear 
attenuator}. The measurement scenario we are considering is 
$\theta\ob = \pi/4$ and $\theta\un = -\pi/8$. We have chosen this scenario, as it gives an 
unbiased impression of how the smoothing technique will perform, i.e., it is not the best nor worst 
measurement scheme for the system but somewhere in between.  Let us choose the system's initial state 
with a mean $\ex{\hbx}_0 = (0,0)\!\tp$ and covariance $V_0 = (\hbar/2)\, {\rm diag}(10,1/2)$. We have 
chosen the initial condition for the covariance so that it is similar to the unconditioned steady-state 
covariance, $V = (\hbar/2)\, {\rm diag}(\infty,1/2)$, while still being finite. 

The trajectories for the $q$ and $p$ quadratures [Figs.~\ref{Fig-Traj}(a) and (b)] show that the smoothed mean 
(red line) seems to be closer, on average, to the true mean (black line) than the filtered mean (blue line). 
Therefore, as expected, the smoothed state provides a better estimate of the true state than the filtered 
state. The \red SWD \blk mean (green line), on the other hand, 
bares very little similarity to the true mean in both quadratures, showing how poorly even the mean of the 
\red SWD \blk state works for this purpose. 

We can also see how the {\red covariances, depicted through the $e^{-1/2}$ contour of the Wigner function corresponding to one standard deviation along the principle axes of the ellipse,} which determine the purity of a Gaussian state, defined as 
$P = (\hbar/2)^N\sqrt{|V|\inv}$, evolve over 
time in Figs.~\ref{Fig-Traj}(c)--(f). At $t = 0$, in Fig.~\ref{Fig-Traj}(c), the filtered, smoothed, and true states 
all begin with the same initial covariance $V_0$. 
As time progresses, the covariances begin to shrink, indicating the increase of the purity, until 
they all reach their steady states at around $t = 0.5T$ in Fig.~\ref{Fig-Traj}(e). At this time the true state 
is guaranteed to be a pure state, and the smoothed state is purer than the filtered state (as the smoothed 
covariance can fit within the filtered covariance). Moreover, at the final time in Fig.~\ref{Fig-Traj}(f), the 
smoothed covariance is exactly the same as the filtered covariance, as expected, since there is no more 
future information to condition on. By contrast, the true state remains in its steady state. 

The covariance of the \red SWD \blk state, as one might expect by 
now, behaves very differently. Initially, the covariance is not the same as that of the initial true state; it is 
substantially smaller. As time progresses, the \red SWD \blk covariance reaches its steady state in 
Fig.~\ref{Fig-Traj}(e), where it is clear that the \red SWD \blk state is unphysical. It has a purity greater than 
unity (the \red SWD \blk covariance can fit entirely within the pure true covariance), violating the \SHUR. At the final 
time, Fig.~\ref{Fig-Traj}(f), the \red SWD \blk covariance matches the filtered state (as well as the smoothed 
state), as it must since there is no future record left.

\section{Optimal Measurement Strategies for Quantum State Smoothing}
\label{sec-Opt}
In the previous section we looked at the improvement in the purity that the smoothed state offered over 
the filtered state. However, the degree of improvement offered by the smoothed state depends on the 
choice of Alice's and Bob's measurements. In this section we study this phenomenon and seek 
a method for predicting the best measurement strategy for Alice and Bob to maximize the purity 
improvement.

In general, the purity of the filtered and smoothed quantum states varies depending on a particular 
realization of the measurement record ${\blk\rm O}$. As a result, it is necessary to average over all possible 
realizations of the observed record ${\blk\rm O}$ in order to draw any conclusions about the purity 
improvement. The measure of purity improvement we will investigate in this paper is the relative average 
purity recovery of a smoothed state. This is the same measure considered in 
Ref.~\cite{CGW19}, given by
\beq
{\cal R} = \frac{{\mathbb E}_{{\blk\rm O}}[P(\rho\sm)] - {\mathbb E}_{{\blk\rm O}}[P(\rho\fil)]}
{{\mathbb E}_{{\blk\rm O}\past{\blk\rm U}}[P(\rho\god)]- {\mathbb E}_{{\blk\rm O}}[P(\rho\fil)]}\,.\label{RAPR}
\eeq
Here ${\mathbb E}_{\blk\rm O}[...]$ ($\mathbb{E}_{{\blk\rm O}\past{\blk\rm U}}[...]$) represents 
averaging over all possible realizations of the observed record ${\blk\rm O}$ (and the past unobserved record 
$\past{\blk\rm U}$), and $P(\rho) = \Tr[\rho^2]$ 
represents the purity of a state $\rho$. The {\blk relative average purity recovery} is a measure of the purity 
increase given from smoothing compared to filtering on average, relative to the maximum average recovery 
possible. 

For Gaussian systems, the expression for the purity recovery can be greatly
simplified. The purity of a Gaussian state is independent of observed and unobserved 
measurement records, and depends solely on the state's covariance matrix. 
Consequently, we only need to consider a relative purity recovery (RPR) \cite{LCW19}, which simplifies the {\blk 
relative average purity recovery} to 
\beq
{\cal R} = \frac{P\sm - P\fil}{P\god - P\fil}\,.
\eeq
Here, for Gaussian states, the purity of the conditioned state is $P\c = (\hbar/2)^N\sqrt{|V\c|\inv}$.
We will now construct three different hypotheses for the optimal measurement scheme for Alice and Bob 
in order to maximize the purity recoveries and compare their predictions to the numerical optimal for the 
physical examples. 

\begin{figure*}[t]
\centering
\includegraphics[width = \textwidth]{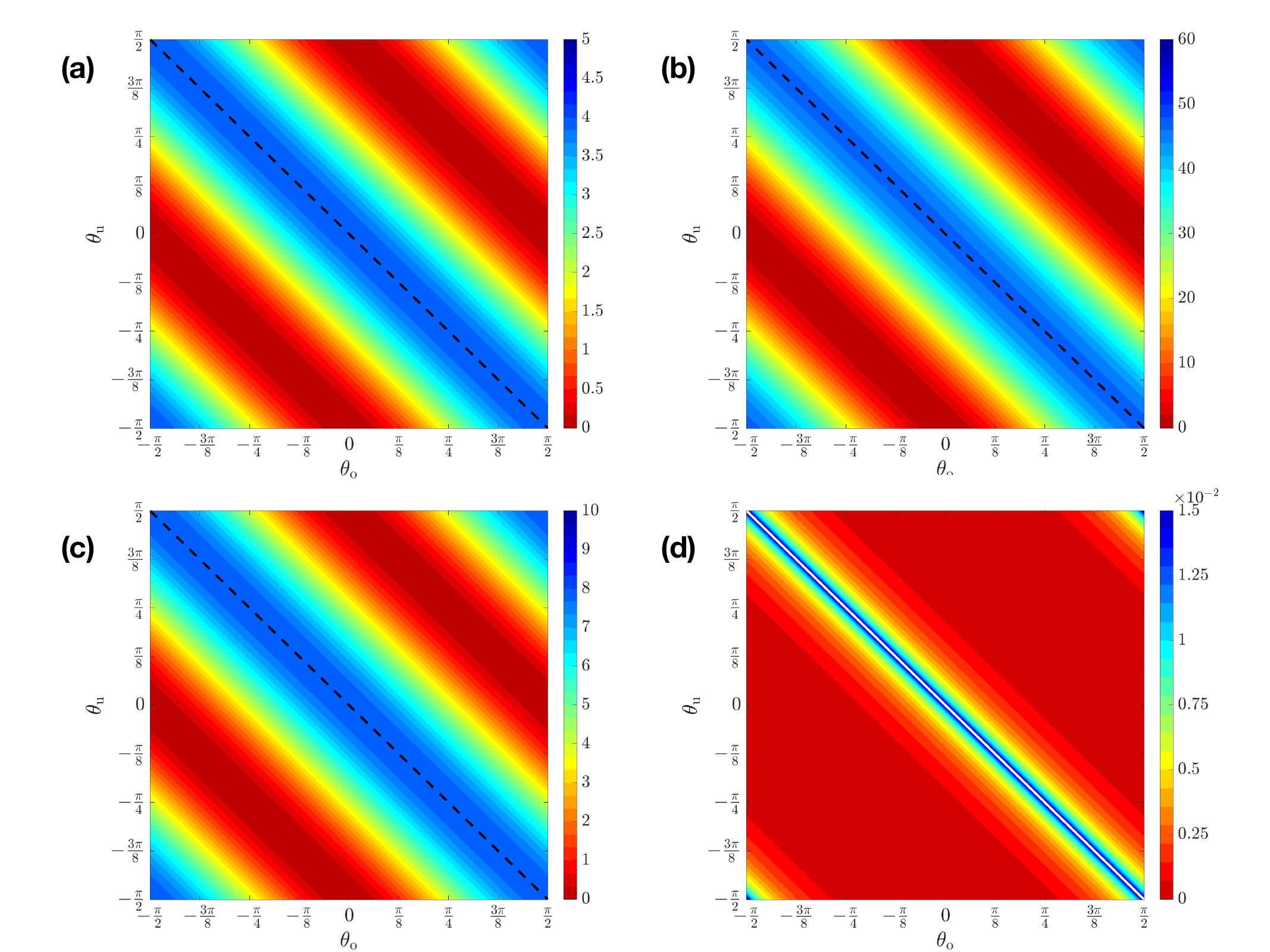}
\caption{Contour plots of {\blk (a) the measurement overlap \erf{Overlap_Meas}, (b) the unobserved 
overlap \erf{Overlap-u}, (c) the observed overlap \erf{Overlap-o} and (d) the RPR for the {\blk noisy linear 
attenuator} system in the
steady state} for different values of the observed (Alice) and unobserved (Bob) homodyne phases. {\blk In 
this example, the range of the unobserved and observed homodyne phases are $\Theta\un = [-\pi/2,\pi/2)$ 
and $\Theta\ob = [-\pi/2,\pi/2)$, respectively.} Note that while (a), (b), and (c) look identical, the scales of the 
contours are very different due to Alice and Bob measuring different channels.    
In (d) we see that the {\blk RPR closely resembles the objective functions in (a)--(c), 
and the optimal RPR (solid white line), obtained numerically, perfectly 
matches the maximum of the objective functions.}
In all plots we consider the case where 
$\gamma\up = 0.999 \gamma\dwn$. Alice perfectly measures the attenuation channel 
($\eta_{\downarrow,{\rm o}} = 1$, $\theta_{\downarrow,{\rm o}} = \theta\ob$), and Bob 
perfectly measures the amplification channel 
($\eta_{\uparrow,{\rm u}} = 1$, $\theta_{\downarrow,{\rm u}} = \theta\un$). We have set $\hbar =2$.}
\label{Fig-SLA}
\end{figure*}

\begin{figure*}
\includegraphics[width = \textwidth]{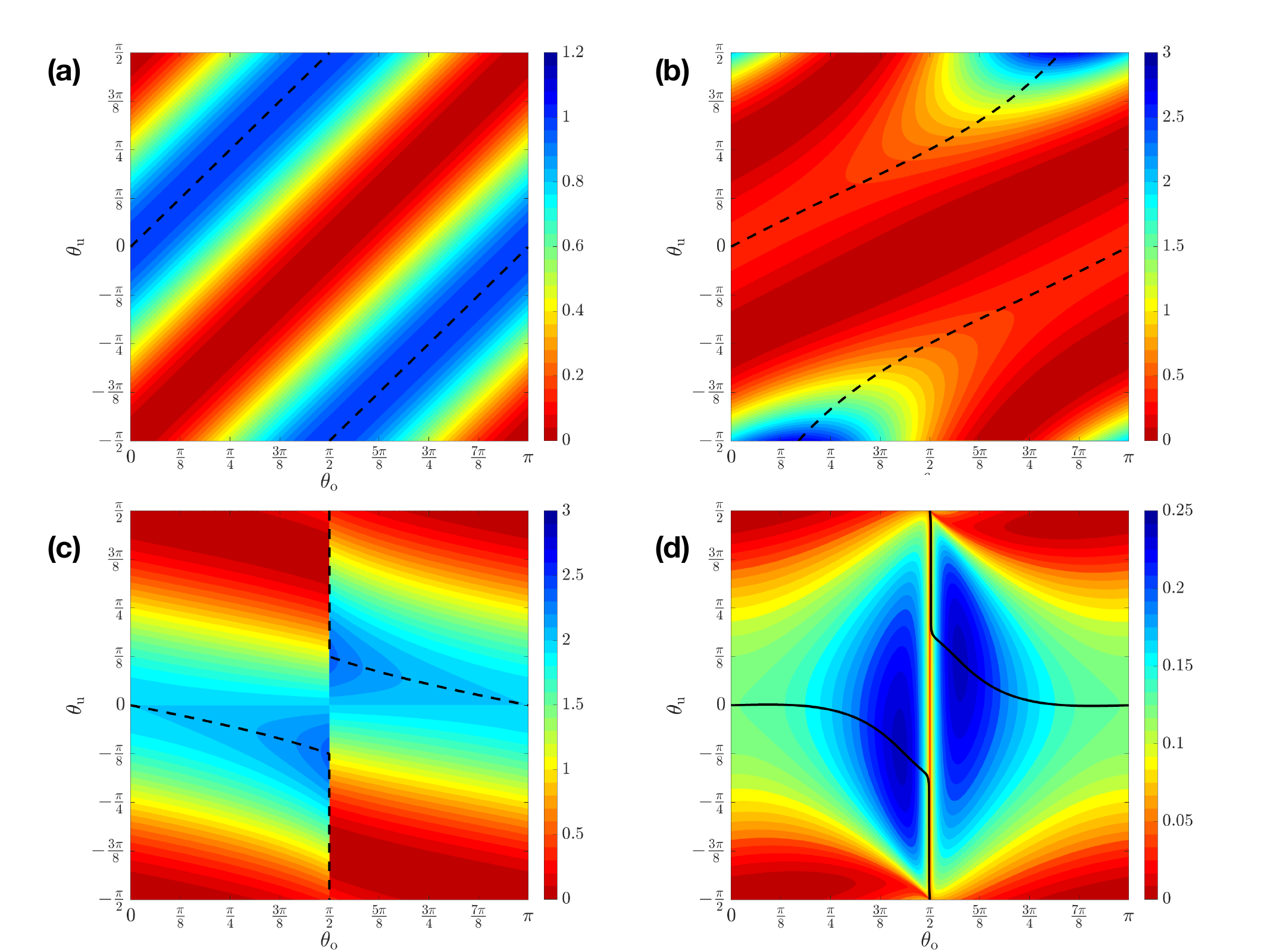}
\caption{Contour plots of (a) the measurement overlap \erf{Overlap_Meas}, (b) the {\blk unobserved overlap 
\erf{Overlap-u}, (c) the observed overlap \erf{Overlap-o}}  and (d) the RPR for the on-threshold OPO in steady 
state for different values of the observed (Alice) and unobserved (Bob) homodyne phases. {\blk In this 
example, the range of the unobserved and observed homodyne phases are $\Theta\un = [-\pi/2,\pi/2)$ and 
$\Theta\ob = [0,\pi)$, respectively.} In (a), we immediately see that the optimal measurement strategy according to
hypothesis A (dashed black line) is very different from the optimal measurement strategy, obtained numerically, 
for RPR [solid black line in (d)], indicating that it is incorrect. In (b), both the solution 
to \erf{guess-2} (dashed black line) and the unobserved overlap behave very differently compared to the 
optimal measurement strategy and the RPR, respectively, in (d). On the contrary, in (c) the solution to 
\erf{guess-1} (dashed black line) gives a close approximation to the optimal measurement strategy. 
Furthermore, the square overlap has developed some of the characteristics of the RPR. In all plots, both 
Alice and Bob measure the same damping channel (with homodyne phases $\theta\ob$ and $\theta\un$, 
respectively) but with $\eta\ob = \eta\un = 0.5$. We have set $\hbar = 2$.}
\label{Fig-OPO}
\end{figure*}

\subsection{Hypothesis A}
The first and simplest guess at the optimal strategy
would be for both Alice and Bob to gather information about the same quantity, e.g.,~both measuring the 
same quadrature. Since in the LGQ case the measurement matrices $C\ob$ and $C\un$ provide 
information about how Alice and Bob measure the system, we can look at the overlap between Alice's and 
Bob's measurement matrices,
\beq\label{Overlap_Meas}
{\cal O}_{\rm m}^{\theta\un}(\theta\ob) = \Tr\left[C\ob^{\theta\ob}(C\un^{\theta\un})\!\tp C\un^{\theta\un}
(C\ob^{\theta\ob})\!\tp\right]\,.
\eeq
{\blk Here, for simplicity, we have used the notation $\theta\ob$ and $\theta\un$ to denote the 
parameters specifying Alice's and Bob's measurement matrices because in this paper we are 
restricting to homodyne measurements of a single channel so that only one angle is needed.  
For the fully general case, we would have to replace $\theta$ by the unraveling matrix $M$ 
as introduced in Sec.~\ref{Mintro}.}

It is easiest to see why we call 
\erf{Overlap_Meas} an overlap function when Alice and Bob only have a single measurement channel at their 
disposal, like in the OPO example presented in Sec.~\ref{Sec-OPO}. In this case, {\blk 
$C\ob$ and $C\un$ become vectors and \erf{Overlap_Meas} is exactly 
the square of their scalar product. This intuition also works for the} {\blk noisy linear attenuator} example where the 
only nonzero element in the resulting matrix corresponds to the squared overlap between Alice's measurement 
on her channel and Bob's measurement on his channel. {\blk Note that the square is important 
here, because there is no difference in the information obtained by a measurement with 
matrix $C$ and one with matrix $-C$, so the objective function ${\cal O}$ should be invariant under a sign 
change.} 

{\blk Thus, for hypothesis A, that Alice should obtain} 
information about the same quantity as Bob, {\blk 
she choose her measurement by} maximizing the measurement overlap function 
\erf{Overlap_Meas} over the allowed range $\Theta\ob$ of homodyne angles. 
That is, she should choose 
\beq\label{guess-1} 
{\blk \theta\ob^\star}(\theta\un) = \arg\max_{\theta\in\Theta\ob} {\cal O}_{\rm m}^{\theta\un}(\theta)\,,
\eeq
where we have written Alice's optimal phase ${\blk \theta\ob^\star}(\theta\un)$ as a function of Bob's homodyne 
phase. In \erf{guess-1} we point out that there is no reason to maximize over Alice's homodyne phase as 
opposed to Bob's homodyne phase as the measurement overlap is identical if $C\ob$ and $C\un$ are 
swapped. 

We test this intuition by considering the two physical systems presented in 
Sec.~\ref{sec-PS} in the steady state. For the {\blk noisy linear attenuator} system, \erf{guess-1} 
results in Alice and Bob measuring their respective channels with homodyne 
phases such that $\theta\ob = -\theta\un$. The negative sign arises from the fact that Alice and Bob 
measure different types of channels, that is, Alice measures an attenuation channel with the 
Lindblad operator $\sqrt{\gamma\dwn}(\hat{q} + i\hat{p})$, and Bob measures the 
amplification channel with the Lindblad operator $\sqrt{\gamma\up}(\hat{q} - i\hat{p})$. 
Comparing the measurement overlap function in Fig.~\ref{Fig-SLA}(a) to the RPR in Fig.~\ref{Fig-SLA}(d), for 
all $\theta\ob = \theta\dwn$ and $\theta\un = \theta\up$, we see that hypothesis A [dashed black line in (a)] 
matches perfectly with the optimal measurement strategy [solid white line in (d)] obtained by a numerical 
search. In fact, the measurement overlap function has a striking resemblance to the RPR for the {\blk noisy linear 
attenuator}.

The {\blk noisy linear attenuator} is, however, a very simple system without any 
unitary dynamics, so we should not jump to any conclusions about hypothesis A's success in predicting the 
optimal measurement. We thus examine the on-threshold OPO system to see how well hypothesis A works. 
Based on \erf{guess-1}, the optimal measurement strategy for the OPO system is $\theta\ob = \theta\un$. This 
is clearly incorrect, as we can see by comparing the measurement overlap function Fig.~\ref{Fig-OPO}(a) to the 
RPR in Fig.~\ref{Fig-OPO}(d). The numerically obtained optimal strategies [solid black lines in (d)] are 
drastically different from the hypothesis $\theta\ob = \theta\un$ [dashed black lines in (a)]. Furthermore, the 
measurement overlap function does not resemble the RPR. Consequently, we have to come up with a more 
refined argument to explain the optimal strategy.

\subsection{Hypothesis B}
{\blk On reflection, it is perhaps not surprising that hypothesis A failed. Alice's ultimate goal is 
to guess Bob's state as well as possible. Why should that be achieved by trying to get the same 
type of information as Bob? Rather, it would seem, Alice should try to get information about 
how Bob's state changes in reaction to his measurement results, which are unknown to her. 
That is, it seems that a better hypothesis would take into account} the 
correlation between the measurement setups and 
the measurement back-action affecting the system. 

We can see how a measurement and its 
corresponding back-action affects the state by comparing the unconditioned equations, 
\erfs{LLE}{UncondV}, to the filtered equations, \erfs{qfm}{qVf}. Specifically, the effect of back-action is 
given by the kick matrix ${\cal K}^+_{\rm r}[V_{\past{\rm R}}]$, 
from which we define a mean-square kick tensor 
\beq
B_{\rm r}^{\theta_{\rm r}} = \K^+_{\rm r}[V^{\theta_{\rm r}}_{\past{\rm R}}]
\K^+_{\rm r}[V^{\theta_{\rm r}}_{\past{\rm R}}]\tp\,.
\eeq 
Here the superscript $\theta_{\rm r}$ specifies the homodyne phase used to calculate {\blk 
the measurement matrix $C_{\rm r}$, the cross-correlation matrix $\Gamma_{\rm r}$, and 
the covariance matrix $V_{\past{\rm R}}$ (which 
all feed into ${\cal K}^+_{\rm r}[V_{\past{\rm R}}]$). The covariance matrix is conditioned on the past 
measurement record $\past{\rm R} = \past{\rm O},\past{\rm U}$, for 
${\rm r} = {\rm o},{\rm u}$ respectively. Note that for ${\rm r} = {\rm u}$ we are considering 
the state conditioned only on Bob's records $\past{\blk\rm U}$, 
with 
a filtered covariance matrix $V^{\theta\un}_{\past{\blk\rm U}}$ 
satisfying }
\beq
\frac{{\rm d}V^{\theta\un}_{\past{\blk\rm U}}}{{\rm d}t} = AV^{\theta\un}_{\past{\blk\rm U}} + 
V^{\theta\un}_{\past{\blk\rm U}} A\tp \!+ D - 
{\cal K}\un^{+} [V^{\theta\un}_{\past{\blk\rm U}}] {\cal K}\un^{+} [V^{\theta\un}_{\past{\blk\rm U}}]\tp\,,
\eeq
similar to \erf{qVf}. 

As Alice is trying to estimate Bob's true state of the system, {\blk the obvious hypothesis 
is} that Alice should choose her 
measurement to observe the back-action (kick) Bob's measurement induces on the system. By choosing this 
measurement scheme, {\blk one would think that} Alice's measurement would contain the most relevant 
information about Bob's measurement {\blk results} and consequently provide a {\blk good} estimate {\blk of} 
the true state. With this in mind, we can construct another {\blk objective} 
function, the unobserved overlap function,
\beq\label{Overlap-u}\blk{
{\cal O}^{\theta\un}\un(\theta) = \Tr\left[C\ob^{\theta} B^{\theta\un}\un(C\ob^{\theta})\!\tp\right]\,,}
\eeq
where we have just replaced Bob's measurement matrix in \erf{Overlap_Meas} with his kick matrix.
{\blk Thus our hypothesis B is that} Alice should choose her measurement in order to maximize the unobserved 
overlap, i.e.,
\beq\label{guess-2}{\blk
{\blk \theta\ob^\star}(\theta\un) = \arg\max_{\theta\in\Theta\ob}{\cal O}\un^{\theta\un}(\theta)\,. }
\eeq

Unsurprisingly, when we consider the {\blk noisy linear attenuator} example, we see in Fig.~\ref{Fig-SLA}(b) that 
the maximum of the unobserved overlap (dashed black line) is obtained when Alice chooses her measurement 
angle such that $\theta\ob = -\theta\un$. {\blk However,} the same cannot be said for the OPO system, as shown 
in Fig.~\ref{Fig-OPO}(b), where both the hypothesized optimal strategy \erf{guess-2} (dashed black line) and the 
unobserved overlap function {\blk bears} little resemblance to the optimal strategy and the RPR in 
Fig.~\ref{Fig-OPO}(d), respectively.

\subsection{Hypothesis C}
Even though hypothesis B {\blk also failed}, the construction is still useful. Specifically, we 
consider the same construction but with Alice and Bob swapped. That is, we consider the counter intuitive 
hypothesis that it is best for Bob to observe {\blk as well as possible} the kick from Alice's 
measurement on the system. Consequently, we define the observed overlap function
\beq\label{Overlap-o}{\blk
{\cal O}\ob^{\theta\ob}(\theta) = \Tr\left[C\un^{\theta}  B\ob^{\theta\ob} (C^{\theta}\un)\!\tp\right]\,,}
\eeq
{\blk where, compared to \erf{Overlap-u}, we have swapped the labels ${\rm o}$ and ${\rm u}$.
With this overlap function defined, our {\blk third and} last hypothesis for the optimal unobserved homodyne 
phase is }
\beq\label{guess-3}{\blk
{\blk \theta\un^\star}(\theta\ob) = \arg\max_{\theta\in\Theta\un} {\cal O}\ob^{\theta\ob}(\theta)\,,}
\eeq
{\blk where we have written Bob's optimal homodyne phase ${\blk \theta\un^\star}(\theta\ob)$ as a function of 
Alice's homodyne phase and $\Theta\un$ is the range of Bob's homodyne phase.}

Once again, when we consider the {\blk noisy linear attenuator}, hypothesis C, 
\erf{guess-3}, still gives the correct optimal solution $\theta\ob = -\theta\un$, as can be seen in 
Fig.~\ref{Fig-SLA}(c). {\blk And} this time when we consider the OPO 
system in Fig.~\ref{Fig-OPO}(c), we {\blk finally do} see  remarkably good agreement between \erf{guess-3} 
(dashed black line) and the optimal measurement strategy [solid black line in Fig.~\ref{Fig-OPO}(d)]. 
Furthermore, the {\blk objective} function for 
hypothesis C {\blk is qualitatively similar} to the RPR, with the distinctive {\blk asymmetrical}
peaks {\blk close to} $\theta\ob = \pi/2$ in Fig.~\ref{Fig-OPO}(d) {\blk appearing also in (c).} 

\begin{figure}
\includegraphics[scale=0.355]{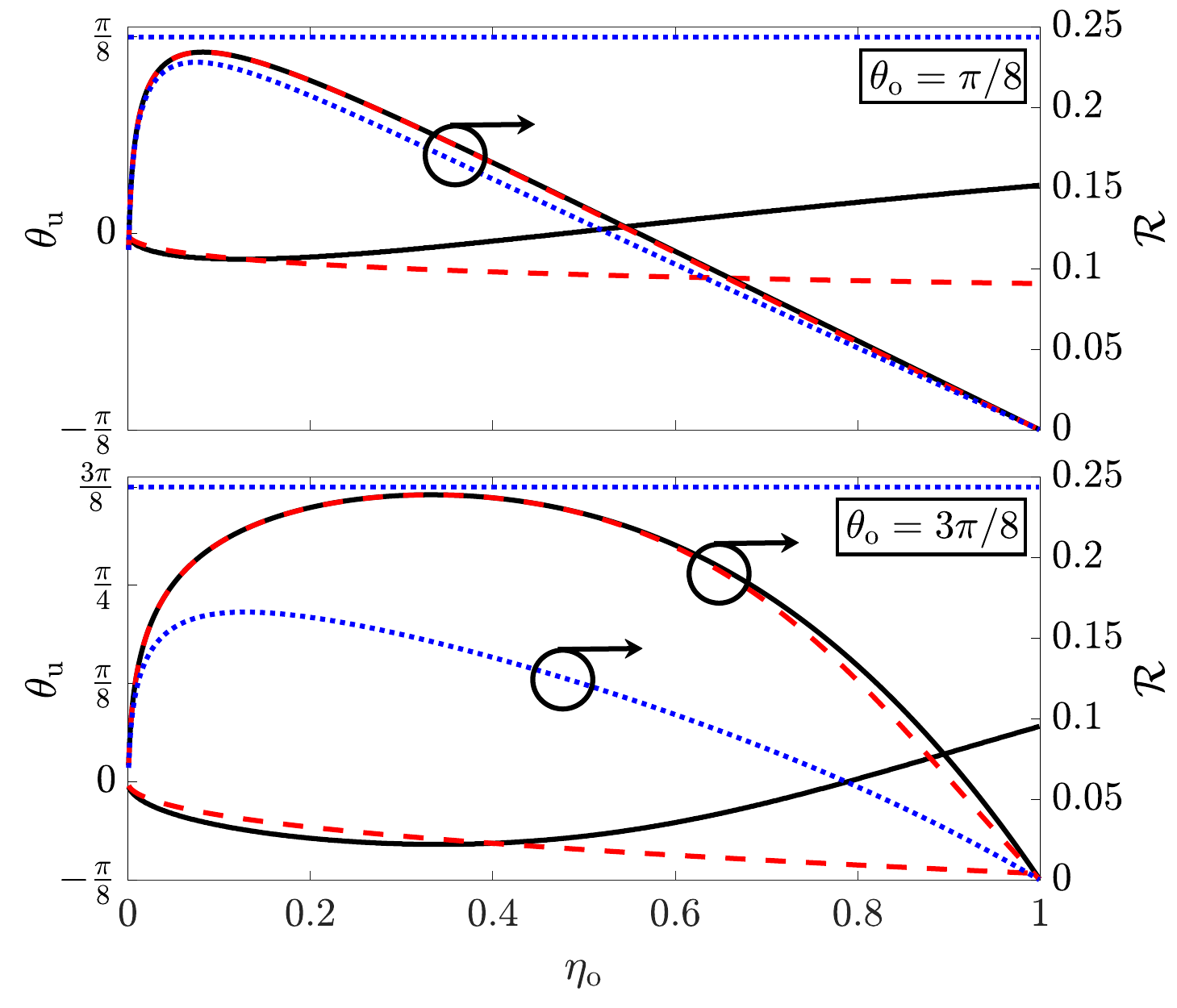}
\caption{The hypothesized and optimal unobserved measurement phases (left-hand-side 
axis) and the RPR (right-hand-side axis) for the OPO system in the steady state with varying observed 
measurement efficiency $\eta\ob$ ($\eta\un = 1-\eta\ob$), for two 
fixed observed measurement phases (top: $\theta\ob = \pi/8$, bottom: $\theta\ob = 3\pi/8$). We consider 
two hypotheses of the optimal measurement strategy for Bob, hypothesis A, \erf{guess-1} (blue dotted 
line), and hypothesis C, \erf{guess-3} (red dashed line), comparing to the numerically obtained optimal 
strategy (black solid line). The results show that the strategy in \erf{guess-3} gives a very close 
approximation to the optimal RPR.}
\label{Fig5}
\end{figure}

The above results were for $\eta\ob = \eta\un$, but we can also check that hypothesis C can 
{\blk reasonably well} predict the optimal measurement strategy for any 
value of measurement efficiencies. We consider the OPO system, choosing two 
measurement phases for Alice ($\theta\ob = \pi/8$ and $3\pi/8$), and compare the optimal measurement 
angle for Bob from the hypotheses and from numerics, for all possible observed measurement 
efficiencies $\eta\ob$ with $\eta\un = 1-\eta\ob$; see Fig.~\ref{Fig5}. 
Comparing the numerically optimal measurement strategy (solid black lines) to hypothesis C 
(dashed red lines), we observe, in both of Alice's measurement phases, that this hypothesis {\blk very well} 
captures the optimal measurement phases $\theta\un$ {\blk when} Alice's efficiency {\blk is low}. At 
higher efficiencies the agreement in optimal phases (see curves associated with the left axis) is not as 
perfect. However, when comparing the 
resulting RPR (curves for the right axis) we observe that the phases given by hypothesis C can still give an 
RPR extremely close to the maximum value. We can also see how well this approximately optimal 
solution does compared to another (suboptimal) measurement strategy, hypothesis A (the blue dotted 
lines), where, especially in the case that $\theta\ob = 3\pi/8$, the differences in the RPR are much larger.

While hypothesis C seems to provide a good approximation of the optimal strategy; it is not based 
on any simple physical intuition, unlike hypothesis A and B. However, {\blk further evidence that
its success here is not a fluke can be gained}  
by applying similar logic to a {\blk very} different type of quantum system, namely, a qubit.

\subsection{Qubit Example}
The single-qubit example we consider in this section is the same as that presented in 
Refs.~\cite{GueWis15,CGW19}. The qubit {\blk has Hamiltonian $\hat{H}_0 = \hbar\omega\s{z}$ and} is 
{\blk coherently} driven {\blk at frequency $\omega$} and is coupled to a bosonic bath. In a frame {\blk that 
removes $\hat{H}_0$}, the master equation for the qubit's unconditioned dynamics is given by
\beq\label{qubit-ME}
\hbar \dot\rho = i[(\Omega/2) \hat\sigma_x,\rho] + \gamma {\cal D}[\hat\sigma_-] \rho\,,
\eeq
where $(\Omega/2) \hat\sigma_x$ is the {\blk driving} Hamiltonian, and 
$\hat\sigma_- \equiv (\hat\sigma_x - i \hat\sigma_y)/2$ is the Lindblad operator. Here $\hat\sigma_k$ are the 
standard Pauli matrices. 
The system-bath coupling rate is denoted by $\gamma$. Alice and Bob could measure the bosonic bath in 
many different ways \cite{WisMil10}. In this work, we only consider homodyne measurements, as we did for
the LGQ systems. {\blk The resulting homodyne photocurrent from monitoring the bath is} 
\beq\label{Photocurrent} 
y_{\rm r}\dd t = \sqrt{\gamma\eta}\, C_{\rm r} \ex{ {\blk \hat{\bf r}}}_{\past{\rm R}}\dd t + \dd w_{\rm r}\,.
\eeq 
Here, {\blk $\hat{\bf r}$ is the 3-vector of Pauli operators 
\beq 
\hat{\bf r} = \left( \s{x}, \s{y} , \s{z}\right)
\tp\,,
\eeq
whose mean is the Bloch vector, which represents the quantum state. 
In \erf{Photocurrent}, this mean is conditioned on the past record 
$\past {\rm R} = \past {\blk\rm O}, \past {\blk\rm U}$ corresponding to $\rm r  = \rm o, \rm u$ respectively.} As 
before, $\eta$ is the measurement efficiency and  the qubit analogue of  measurement matrix is 
\beq \label{defCqubit}
C_{\rm r} = [\cos(\theta_{\rm r}), \sin(\theta_{\rm r}), 0]
\eeq
 for this particular example. 

We will restrict our analysis to two cases for the measurement: $x$ homodyne and $y$ homodyne, i.e., {\blk 
$\theta_{\rm r} =  0$ and $\theta_{\rm r} = \pi/2$, respectively. These choices are the natural ones }
given the symmetries of \erf{qubit-ME}. These are 
named {\blk $x$  and $y$ homodyne} because of the corresponding Pauli {\blk operator} appearing in the 
mean photocurrent signal, from \erf{defCqubit}.  These {\blk two cases} best illuminate the 
effect of measurement choices on the {\blk relative average purity recovery} in the limit of large $\Omega$. Here 
we choose $\Omega = 5 \gamma$. We will also assume that Alice and Bob monitor this bath with 
equal measurement efficiencies, i.e., $\eta\ob = \eta\un = 1/2$. We follow the analysis of the qubit's {\blk relative 
average purity recovery} presented in Ref.~\cite{CGW19}, using numerical analyses, because there is no closed-
form solution for the qubit case. 

\begin{figure}
\includegraphics[width=8.4cm]{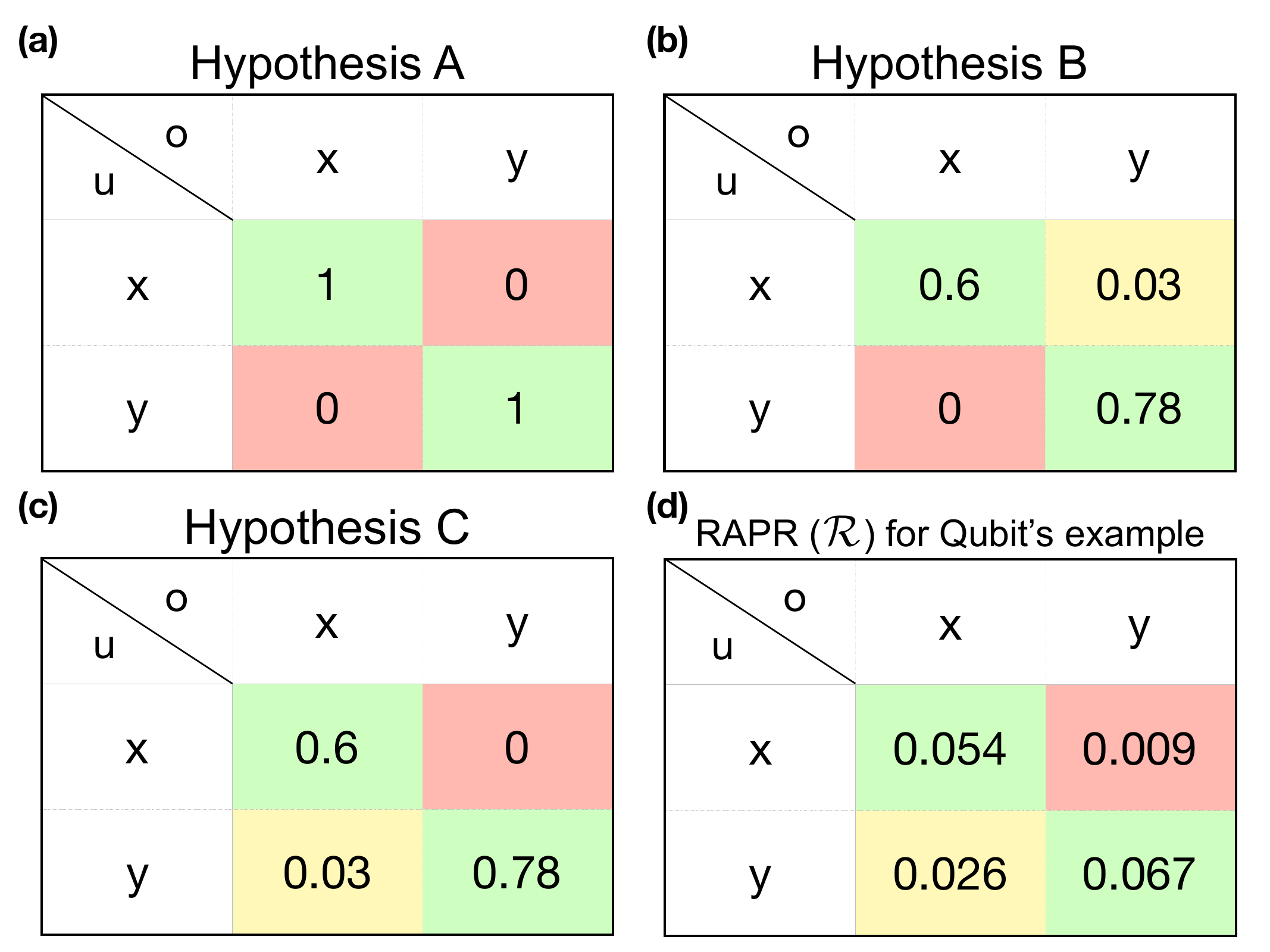}
\caption{Analysis of hypothesis A, B, and C and the {\blk relative average purity recovery} (${\cal R}$) for the 
example of a driven {\blk qubit coupled dissipatively} to 
a bosonic bath. We {\blk restrict Alice and Bob to only two} measurement choices, 
either $x$  or $y$ homodyne. The numerical values in tables (a), (b), and (c) are the 
{\blk objective functions for the respective hypotheses. 
For B and C this required stochastic simulated, and we used} 3000 records each. The qubit's {\blk relative average 
purity recovery} [Table (d)] is obtained using the numerical 
techniques presented in Ref.~\cite{CGW19}, simulating $3000$ observed and $10000$ unobserved records for 
both measurement settings. Here the coloured cells indicate good (green), moderate (yellow), and bad (red) 
improvement. {\blk Only hypothesis C [Table (c)] correctly predicts the pattern of the {\blk relative average purity 
recovery}.}}
\label{Fig6}
\end{figure}

By numerically generating a large ensemble of measurement records and qubit trajectories (including
true states, filtered states, and smoothed states) as functions of time, we can calculate the purity recovery 
averaged over the observed 
records as in Eq.~\eqref{RAPR}. Since we are interested in the steady-state regime, we need
to consider the time period in the simulation to study the qubit's dynamics independently of the transient 
effects at the start and end of the interval. 
Using the dephasing time defined as $T_\gamma = 1/\gamma$ and the final time 
$T = 8 T_\gamma$, we choose the steady-state period to be $\mathfrak{T}_{\rm ss} = [ 4.5 T_\gamma, 6 
T_\gamma]$. We show in Fig.~\ref{Fig6}{\blk (d)}, the $2\times 2$ table of the {\blk relative average purity 
recovery} averaged over the steady-state period quoted from Ref.~\cite{CGW19}, considering four options of 
Alice's ({\blk\rm O}) and Bob's ({\blk\rm U}) measurements.
The combination with the best performance is when Alice and Bob measure the same quadrature and the worst 
performance when Alice measures the $y$ quadrature and Bob measures the $x$ quadrature. Thus we next ask 
whether hypothesis A, B, or C can correctly predict all features of the {\blk relative average purity recovery}.

As we have already defined the measurement matrix for this qubit example, $C_{\rm r}$ in \erf{Photocurrent}, the 
measurement overlap and optimal measurement strategy for hypothesis A are as defined in 
\erf{Overlap_Meas} and (\ref{guess-1}), respectively. As 
we are only considering two measurement possibilities for Alice and Bob, the maximization over the range of 
the unobserved homodyne phases can be replaced by maximizing over the set $\Theta\ob = \{0,\pi/2\}$. 
Calculating the measurement overlap for {\blk the} four possible measurement combinations for Alice and Bob, we 
see, in Fig.~\ref{Fig6}(a), that the optimal measurement strategy, according to hypothesis A, occurs when 
Alice and Bob choose the same measurement. This is consistent with the greatest improvement in the average 
purity of the smoothed state, as seen in Fig.~\ref{Fig6}(d). However, in the cases where Alice and Bob choose 
different measurements, we see that the measurement overlap function suggest that there is no difference 
between these last two cases, which clearly is not true when we look at the {\blk relative average purity recovery}. 
Once again, hypothesis A is {\blk not very accurate.}

To analyze hypotheses B and C for the qubit case, we need to 
define a quantity that resembles the {\rm mean-square kick tensor} of the LGQ system. The {\blk kick 
matrix} is defined in 
Eqs.~\eqref{qfm} and \eqref{truest} and describes the measurement back-action for an LGQ system in terms 
of the change in the system's expectation values in the $q$ and $p$ quadratures. Given a measurement 
setting $\rm r \in \{ \rm o, \rm u \}$ and its corresponding measurement record 
$\past{\blk\rm R} \in \{ \past {\blk\rm O}, \past {\blk\rm U} \}$, respectively, we can {\blk rewrite} the {\blk mean-
square kick tensor} as 
\begin{align}\label{dir-Kick}
{\blk B_{\rm r}}\dd t  = \K^{+}_{\rm r} [V_{\past{\blk\rm R}} ]\K^{+}_{\rm r} [V_{\past{\blk\rm R}} ]\tp \dd t = 
\mathbb{E}_{\past{\blk\rm R}} \left [{\rm d}\ex{\hbx}_{\past{\blk\rm R}} {\rm d}{\ex{\hbx}\tp_{\past{\blk\rm R}}} 
\right].
\end{align}
Here $\ex{\hbx}_{\past{\blk\rm R}}$ is the {\blk LGQ phase-space} mean conditioned on a realization of the (past) 
record $\past{\blk\rm R}$, and the expected average on the right-hand side of \erf{dir-Kick} is over all possible 
record realizations. The right-hand side is exactly the mean-square change (during an infinitesimal time 
${\rm d}t$) of the system's expectation values, {\blk in a tensorial sense}, averaging over all the possible records. 
Therefore we can define an analogous quantity to the mean-square kick tensor for the qubit system as
\beq
\begin{split}\label{qbkick}
{\blk B_{\rm r}}= \mathbb{E}_{\past{\blk\rm R}}\left\{ \frac{1}{|\mathfrak{T}_{\rm ss}|} \,\!\! \sum_{t \in 
\mathfrak{T}_{\rm ss}} \left[ \dd\ex{{\blk \hat{\bf r}}}_{\past{\rm R}}(t) \, \dd\ex{{\blk \hat{\bf r}}}_{\past{\rm R}}(t)\tp \right]\right\}\,,
\end{split}
\eeq
for the steady-state period $\mathfrak{T}_{\rm ss}$ of length $|\mathfrak{T}_{\rm ss}|$. 

Now that we have defined the mean-square kick tensor for the qubit setting, we 
can formalize and analyze both hypothesis B and C. {\blk We will begin with hypothesis B, where the 
unobserved overlap and optimal measurement strategy are as defined in \erfs{Overlap-u}{guess-2}, where, 
as in hypothesis A, we maximize over the set $\Theta\ob = \{0,\pi/2\}$.}
As seen from the four possible measurement combinations for Alice and Bob in Fig.~\ref{Fig6}(b), the 
optimal measurement choice for Alice, according to \erf{guess-2}, occurs when Alice and Bob choose the 
same measurement, {\blk and best of all is when both choose homodyne measurements along the $y$ direction. 
This is consistent with the actual} {\blk relative average purity recovery}, as seen in 
Fig.~\ref{Fig6}(d). However, when we investigate the other measurement combinations, specifically when 
Alice and Bob choose different measurements, we see that the unobserved overlap function does not 
reproduce the {\blk pattern seen for the {\blk relative average purity recovery}. That is, it predicts that smoothing 
would be better if Alice chose 
$y$ and Bob $x$ rather than the other way around, whereas the truth is the opposite.}

For hypothesis C, the roles of Alice and Bob are reversed compared to hypothesis B, and the optimal 
measurement for Bob is given by \erf{guess-3} {\blk and the observed overlap defined in \erf{Overlap-o}. As 
was the case for hypothesis B, we are restricting our analysis to two measurement choices for Alice and Bob, 
and the maximization is instead over the set $\Theta\un = \{0,\pi/2\}$.}
{\blk For the four possible measurement choices for Alice and Bob}, shown in Fig.~\ref{Fig6}(c), 
{\blk the best combination is when both measure $y$ and the second best when both measure $x$, 
consistent with the {\blk relative average purity recovery}, Fig.~\ref{Fig6}(d), and the same as in hypothesis B. 
However, unlike for hypothesis B, this time the objective} 
function for the cases when Alice and Bob choose different measurements also {\blk matches} the {\blk relative 
average purity recovery}. 
{\blk This shows that hypothesis C is better at predicting when smoothing will work well than either 
hypothesis A or hypothesis B.} This is consistent with the results obtained for the LGQ systems. 

\section{Conclusion}
In this paper we provided a detailed derivation of the smoothed quantum state 
for LGQ systems 
{\blk and contrasted it with the theory of the smoothed weak-value state.} To 
exemplify the differences between these techniques, we simulated a single 
trajectory and witnessed clear differences in the dynamics of the estimates by looking at the filtered, the 
\red SWD \blk {\blk state}, and the smoothed quantum states for LGQ systems. 
{\blk As expected, the last of these provides the best estimate of the 
true state conditioned on the results of measurements on a channel unavailable to the 
observer, Alice, as well as on the results of Alice's measurements.} 

{\blk A key question of interest is how much improvement smoothing can offer 
relative to filtering and how this depends on the measurement choices of 
Alice and Bob (the observer of the channel unavailable to Alice). 
We studied this through the purity recovery of smoothing over filtering 
relative to the maximum possible purity recovery.  
We constructed three different hypotheses about what properties of Alice and Bob's 
measurements would lead to higher relative purity recovery. 

We found that the only hypothesis that worked, qualitatively, for the two LQG 
systems we studied is the most counter intuitive of the three.  
It is the hypothesis that says Bob should choose his measurement so that his signal tells 
him as much as possible about the {\em disturbance} to the state caused by Alice's  
measurements. This is counter intuitive because one would have thought that it is 
Alice, the one doing the smoothing, who needs to be able to infer as accurately as 
possible the disturbance to the state caused by Bob's measurement. After all, it is 
the existence of this disturbance that makes Alice's filtered state impure and allows 
the possibility of increasing the purity by smoothing.}


{\blk The qualitative success of our third hypothesis is the 
main result of this paper. However, it presents a puzzle because 
it is not grounded in physical intuition. 
For this reason we also put our three hypotheses to the test 
on a very different system, specifically, a qubit system, not an LGQ system. 
We formulated the problem in a closely analogous way to that used for LGQ systems
and found that, once again, our third hypothesis was clearly superior to the other two 
in predicting which combinations of measurements by Alice and Bob would give 
better relative purity recovery than the other combinations.  


It can be hoped that further study will elucidate 
why it is preferable for Bob to measure the system so as to detect the 
`kick' to the state by Alice's measurement, rather than the converse.  
Another interesting question is what would happen to the smoothed state if Alice were to assume the incorrect 
 type of measurement for Bob.}
{\blk Could the smoothed state be a {\em worse} estimate of the true state than the filtered state? 
The LGQ formalism offers a convenient way to explore this because of the possibility 
of semianalytic solutions.} 
There is also a great deal of work to be done in comparing the 
various other ways of utilising past and future measurement information, such as the most likely path 
formalism \cite{Cha13, Web14}, and in applying these theories to the LGQ scenario.  

\begin{acknowledgments}
We would like to thank Prahlad Warszawski for useful discussions regarding the 
retrofiltered effect. We acknowledge the traditional owners of the land on which this work was undertaken 
at Griffith University, the Yuggera people. This research is funded by the Australian Research Council 
Centre of Excellence Program through Grant No. CE170100012.  A.C.~acknowledges the support of the Griffith University 
Postdoctoral Fellowship scheme.
\end{acknowledgments}


%

\end{document}